\documentclass[english]{aa}
\usepackage{mathptmx}
\usepackage[T1]{fontenc}
\usepackage[latin9]{inputenc}
\usepackage{array}
\usepackage{url}
\usepackage{amstext}
\usepackage{amssymb}
\usepackage{wasysym}
\usepackage{graphicx}
\usepackage[authoryear]{natbib}

\makeatletter

\newcommand{\noun}[1]{\textsc{#1}}
\newcommand{\lyxmathsym}[1]{\ifmmode\begingroup\def\b@ld{bold}
  \text{\ifx\math@version\b@ld\bfseries\fi#1}\endgroup\else#1\fi}

\providecommand{\tabularnewline}{\\}

\newenvironment{lyxlist}[1]
{\begin{list}{}
{\settowidth{\labelwidth}{#1}
 \setlength{\leftmargin}{\labelwidth}
 \addtolength{\leftmargin}{\labelsep}
 }}
{\end{list}}

\makeatother

\usepackage{babel}
\begin{document}
\abstract{}{In this paper we present a classification of emission-line
galaxies at intermediate and high redshifts ($0.5\lesssim z\lesssim1.0$
for optical spectra, $z\gtrsim2.5$ for near-infrared spectra), using
the $D_{n}(4000)$ index as a supplementary diagnostic. Our goal is
to complement the diagnostic based only on emission-line ratios from
the blue part of the spectra, which suffer from some limitations for
the classification of Seyfert 2 and composite galaxies.}{We used
a sample of 89~379 galaxies with a good signal-to-noise ratio from
the Sloan Digital Sky Survey (data release 7). Using the classification
scheme presented in Paper I, we classified these galaxies with a diagnostic
diagram involving the {[}\noun{Oiii}{]}$\lambda5007$/H$\beta$ and
{[}O\noun{ii}{]}$\lambda\lambda3726+3729$/H$\beta$ emission-line
ratios. Then we derived a supplementary diagnostic involving $D_{n}(4000)$
to improve this classification, in the regions where objects of different
types are mixed. To show the validity of our spectral classification
we established success-rate and contamination charts, then we compared
our results to those obtained with the reference classification that
was scheme obtained also using H$\alpha$, {[}N\noun{ii}{]}$\lambda$6584,
and {[}S\noun{ii}{]}$\lambda\lambda$6717+6731 emission lines. }{We
show that our supplementary classification based on the $D_{n}(4000)$
index allows to separate unambiguously star-forming galaxies from
Seyfert 2 in the region where they were mixed in Paper~I. It also
significantly reduces the region where star-forming galaxies are mixed
with composites.}{}


\title{Spectral classification of emission-line galaxies from the Sloan
Digital Sky Survey}

\subtitle{II. A supplementary diagnostic for AGNs\\
using the $D_{n}(4000)$ index}

\author{J. Marocco\and E. Hache\and F. Lamareille}

\institute{Laboratoire d\textquoteright{}Astrophysique de Toulouse-Tarbes, Université
de Toulouse, CNRS, 14 avenue Edouard Belin, F-31400 Toulouse, France\\
\email{flamare@ast.obs-mip.fr}}

\date{Received ; Accepted}

\maketitle

\section{Introduction\label{sec:Introduction}}

There are several existing types of emission-line galaxies: the two
main classes are star-forming galaxies (hereafter SFG) and active
galactic nuclei (hereafter AGN). Emission lines are observed in star-forming
galaxies because gas is ionized by new hot stars. In contrast, AGN
galaxies contain a supermassive blackhole, and their emission lines
come from gas ionization by the light emitted from their accretion
disk. AGN can be classified in several types, but we only consider
narrow-line AGNs, which can be confused with SFG, i.e. Seyfert 2 galaxies
and LINERs (low-ionization nuclear emission-line region). We do not
consider Seyfert 1 galaxies because they can be easily distinguished
from SFGs by their wide Balmer emission lines. A third class of emission-line
galaxies is what we call {}``composites''. Composites show emission
lines which are due both to recent star formation and to an AGN.

To classify emission-line galaxies, one may use two diagnostic diagrams
depending on the redshift range: the first one is known as the BPT
diagnostic \citep{1981PASP...93....5B}, later studied by \citet{2001ApJ...556..121K}
who used it to separate AGN from SFG thanks to theoretical models.
\citet{2003MNRAS.346.1055K} revised Kewley's work and allowed going
deeper into the classification process by showing a third type of
galaxies called composites. It was then again revised by \citet{2006MNRAS.372..961K},
who improved the classification of AGNs into Seyfert 2 and LINERs.
This diagnostic uses log$\left([\mathrm{O\textrm{\textsc{iii}}}]\lambda5007/\mathrm{H}\beta\right)$
vs. log$\left([\mathrm{N}\textsc{ii}]\lambda6583/\mathrm{H}\alpha\right)$
and log$\left([\mathrm{O\textrm{\textsc{iii}}}]\lambda5007/\mathrm{H}\beta\right)$
vs. log$\left([\mathrm{S}\textsc{ii}]\lambda\lambda6717+6731/\mathrm{H}\alpha\right)$
diagrams and may be used up to $z\apprle0.5$ with optical spectrographs.
Other diagnostics have been used in the past in the same diagrams
\citep{1980A&A....87..152H,1987ApJS...63..295V,1997ApJS..112..315H}.
We use \citet{2006MNRAS.372..961K} as a reference since it is the
latest widely used diagnostic and is based on the biggest sample.
We refer the reader to \citet{2006MNRAS.372..961K}, \citet{2006ApJ...650..727C},
and references herein for comparisons of these diagnostics. See also
\citet{2006MNRAS.371.1559G} for a specific discussion on low-metallicity
AGNs.

The second diagnostic was originally proposed by \citet{1996MNRAS.281..847T}
and studied later by \citet{1997MNRAS.289..419R}. This diagnostic
is useful at intermediate and high redshift when some emission lines
used in the BPT diagnostic are no longer observed by getting red-shifted
out of spectrographs. \citet[hereafter L04]{2004MNRAS.350..396L}
established a classification using empirical demarcation lines in
the diagnostic diagram showing $\log\left([\mathrm{O}\textsc{iii}]\lambda5007/\mathrm{H}\beta\right)$
vs. $\log\left([\mathrm{O}\textrm{\textsc{ii}}]\lambda\lambda3726+3729/\mathrm{H}\beta\right)$,
which may be used up to $z\apprle1.0$ with optical spectrographs,
or even at $z\gtrsim2.5$ with near-infrared spectrographs(where optical
diagnostics cannot be used). In Paper~I \citep{2009lama}, one of
us proposed revised equations for the classification that we use in
this paper. We know that the \citet[hereafter L10]{2009lama} diagnostic
implies a loss of Seyfert 2 galaxies, because of the region where
Seyfert 2 and SFGs get mixed. As discussed in Paper~I, the L10 diagnostic
also cannot unambiguously separate composites from SFGs or LINERs.
The goal of this paper is to try to solve these two limitations with
a different approach. Following the idea under the {}``DEW'' diagnostic
introduced by \citet{2006MNRAS.371..972S}, we use the $D_{n}(4000)$
index to derive a supplementary diagnostic. \citet{2011ApJ...728...38Y}
have already derived a similar new diagnostic based on $U-B$ rest-frame
colors. Compared to the present paper, it does suffer from the following
limitations. It is based on rest-frame colors whose calculation may
suffer from biases from imperfect $k$-correction at high redshift
(unless such colors are integrated directly from the spectra), it
does not provide a distinction between Seyfert 2 galaxies and LINERs,
and it does not provide a way to isolate at least a fraction of composite
galaxies. Conversely, that diagnostic has the advantage of only relying
on the detection of $[\mathrm{O}\textsc{iii}]\lambda5007$ and $\mathrm{H}\beta$
emission lines.

Our goal is to provide a diagnostic that can be used to classify intermediate-
or high-redshift emission-line galaxies as closely as possible to
local universe studies. The older L04 diagnostic has already been
used in various studies, such as star formation rates \citep{2009ApJ...694.1099M},
metallicities \citep{2006MNRAS.369..891M,2006A&A...448..907L}, AGN
populations \citep{2010A&A...510A..56B}, gamma ray burst hosts \citep{2009ApJ...691..182S},
and clusters \citep{2009MNRAS.398..133L}. Results provided in Paper~I
and here may be used to revise spectral classification of emission-line
galaxies in intermediate redshift optical galaxy redshift surveys
such as VVDS \citep{2005A&A...439..845L,2008A&A...486..683G}, zCOSMOS
\citep{2009ApJS..184..218L}, DEEP2 \citep{2003SPIE.4834..161D},
GDDS \citep{2004AJ....127.2455A}, GOODS \citep{2010A&A...512A..12B},
and others. We hope it will also serve as a reference for ongoing
or future high-redshift surveys involving future spectrographs: in
the optical, MUSE on VLT \citep{2010SPIE.7735E...7B} or DIORAMAS
on EELT \citep{2010SPIE.7735E..75L}; or in near-infrared (at $z\gtrsim2.5$),
EMIR on GTC \citep{2006SPIE.6269E..40G,2005RMxAC..24..154C}, KMOS
on VLT \citep{2006SPIE.6269E..44S}, MOSFIRE on Keck \citep{2008SPIE.7014E..99M}. 

It is worth mentioning here as a warning that \citet{2008MNRAS.391L..29S}
demonstrate that a fraction -- whose value is still uncertain -- of
the galaxies classified as LINERs or composites by emission-line diagnostics
may be actually {}``retired'' galaxies. Ionization in such galaxies
would be produced by post-AGB stars and white dwarfs. The reader should
therefore be aware that galaxies that we refer to as LINERs or composites
\emph{might not} contain an AGN. \citet{2011MNRAS.tmp..249C} have
derived a diagnostic that isolates this class of {}``retired'' galaxies.
This diagnostic is based on H$\alpha$ and {[}N\noun{ii}{]}$\lambda$6583
emission lines. It is, however, beyond the scope of this paper to
derive a similar diagnostic that can be used on higher redshift spectra,
but it may be the goal of a future work.

This paper is organized as follows. We first present the data and
how we selected them (Sect.~\ref{sec:Data-selection}), then we summarize
of existing classification schemes (Sect.~\ref{sec:Existing-classification-schemes}).
In Sect.~\ref{sec: Limits of blue classifications}, we discuss the
limits of the L10 and DEW diagnostics. Finally we present our supplementary
diagnostic in Sect.~\ref{sec:Spectral-classification}.

\section{Data selection\label{sec:Data-selection}}

We used a sample of 868~492 galaxies from the SDSS (Data Release
7, \citealp{2009ApJS..182..543A}, available at: \url{http://www.mpa-garching.mpg.de/SDSS/DR7/})
with redshifts between 0.01 and 0.3. Actually the sample originally
contained measurements for 927~552 different galaxies, but there
are 109~219 duplicate spectra (twice or more), so we averaged these
duplicated measurements in order to increase the signal-to-noise ratio,
and filtered out those that do not increase the averaged signal-to-noise
ratio. Among others, these data contain measurements of the equivalent
widths of the following emission lines: {[}O\noun{iii}{]}$\lambda$5007,
{[}O\noun{ii}{]}$\lambda\lambda$3726+3729, {[}N\noun{ii}{]}$\lambda$6583,
{[}\noun{Sii}{]}$\lambda\lambda$6717+6731, H$\beta$, H$\alpha$,
and {[}NeIII{]}$\lambda$3969. Balmer emission-line measurements were
automatically corrected for any underlying absorption. The spectral
coverage of SDSS is 3800-9200~$\lyxmathsym{\AA}$, and the mean resolution
of the spectra $1800\lesssim\lambda/\triangle\lambda\lesssim2200$.
We also retrieved the value of the $D_{n}(4000)$ index \citep{1999ApJ...527...54B},
which were measured on emission-line subtracted spectra.

We filtered our data for a specific signal-to-noise ratio (in equivalent
width) greater than five in order to keep the same selection as in
Paper~I, also eliminating data with positive equivalent width, which
would involve absorption lines. We did not apply this selection to
the {[}NeIII{]}$\lambda$3969, which is only used as an optional measurement
in the DEW diagnostic. This finally leads to 89~379 galaxies. All
classifications and plots presented in this paper were processed by
the {}``JClassif'' software, part of the {}``Galaxie'' pipeline
available at: \url{http://www.ast.obs-mip.fr/galaxie/}. 

Throughout this paper, as in Paper~I, all emission line ratios are
\emph{equivalent width} ratios rather than flux ratios. This is done
to eliminate any dependence that may exist (mainly for the $[\mathrm{O}\textsc{ii}]/\mathrm{H}\beta$
emission line ratio) between the derived diagnostic and the dust properties
of the sample. Indeed, equivalent width ratios are sensitive not to
dust attenuation, but only to the ratio between continuum fluxes below
each lines. Considering {[}O\noun{ii}{]} and H$\beta$, this parameter
should not evolve strongly between galaxies with similar properties
in the diagnostic diagrams, even if they are at different redshifts,
keeping the consistency of the diagnostics \citep[see also][]{2006A&A...448..893L,2009A&A...495...73P}.

\section{Existing classification schemes\label{sec:Existing-classification-schemes}}

\subsection{K06 diagnostic}

As in Paper~I, we use the a simplified version of the diagnostic
from \citet[hereafter K06]{2006MNRAS.372..961K} as the reference
classification. In the two main K06 diagrams, we use the following
demarcation lines:

\begin{equation}
\log\left([\mathrm{O}\textsc{iii}]/\mathrm{H}\beta\right)=0.61/[\log\left([\mathrm{N}\textsc{ii}]/\mathrm{H}\alpha\right)-0.05]+1.30,\label{eq: Kauffmann red}
\end{equation}
where AGNs are above this curve, and

\begin{equation}
\log\left([\mathrm{O}\textsc{iii}]/\mathrm{H}\beta\right)=0.61/[\log\left([\mathrm{N}\textsc{ii}]/\mathrm{H}\alpha\right)-0.47]+1.19,\label{eq: Kewley red}
\end{equation}
with SFGs below this curve. Composites fall between these two curves.
Moreover, AGNs can be subclassified into Seyfert 2 and LINERs using
the line

\begin{equation}
\log\left([\mathrm{O}\textsc{iii}]/\mathrm{H}\beta\right)=1.89\times\log\left([\mathrm{S}\textrm{\textsc{ii}}]/\mathrm{H}\alpha\right)+0.76.\label{eq: Kewley LINERs/SFG}
\end{equation}
Seyferts 2 are above this line, LINERs are below. 

Our K06 diagnostic is simplified since it does not use the last log$\left([\mathrm{O\textrm{\textsc{iii}}}]\lambda5007/\mathrm{H}\beta\right)$
vs. log$\left([\mathrm{O}\textsc{i}]\lambda6300/\mathrm{H}\alpha\right)$
diagram. This is a reasonable approximation since the $[\mathrm{O}\textsc{i}]\lambda6300$
emission line is weaker than the others, hence not detected in most
intermediate- and high-redshift spectra where the signal-to-noise
ratio is typically low.

\subsection{L10 diagnostic\label{sub:L10-classification}}

The L10 diagnostic has been defined in Paper~I. We summarize here
the main equations, but we refer the reader to Paper~I for details.
The first equation separates SFGs from AGNs: 

\begin{equation}
\log\left([\mathrm{O}\textsc{iii}]/\mathrm{H}\beta\right)=0.11/[\log\left([\mathrm{O}\textsc{ii}]/\mathrm{H}\beta\right)-0.92]+0.85,\label{eq: New blue SFG/AGN}
\end{equation}
where AGNs are above this curve. The second equation separates Seyfert
2 from LINERs in the AGN region:
\begin{equation}
\log\left([\mathrm{O}\textsc{iii}]/\mathrm{H}\beta\right)=0.95\times\log\left([\mathrm{O}\textsc{ii}]/\mathrm{H}\beta\right)-0.40.\label{eq: New blue LINERs/Sey2}
\end{equation}
Seyferts 2 are above this line. Then, we define a region where some
Seyfert 2 (26\% of them) are mixed with a majority of SFGs (21.5\%
contamination by Seyfert 2). This region, called {}``SFG/Sy2'',
is located below Eq.~\ref{eq: New blue SFG/AGN} and above the line

\begin{equation}
\log\left([\mathrm{O}\textsc{iii}]/\mathrm{H}\beta\right)=0.30.\label{eq: New blue SFG-Sey2 mix}
\end{equation}
Finally, we define the region where most of the composites fall (64\%
of them), even if this region is dominated by SFGs (79\%) and also
contains some LINERs (2\%). This region, called {}``SFG-LIN/comp'',
can be located by the two following inequalities: 

\begin{equation}
\log\left([\mathrm{O}\textsc{iii}]/\mathrm{H}\beta\right)\le-(x-1)^{2}-0.1x+0.25,\label{eq: New blue composite 1}
\end{equation}

\begin{equation}
\log\left([\mathrm{O}\textsc{iii}]/\mathrm{H}\beta\right)\ge(x-0.2)^{2}-0.60,\label{eq: New blue composite 2}
\end{equation}
with $x=\log\left([\mathrm{O}\textsc{ii}]/\mathrm{H}\beta\right)$.
Unlike in Paper~I, we now divide the {}``SFG-LIN/comp'' region
for clarity into {}``SFG/comp'' and {}``LIN/comp'' regions, and
the separation between SFG and LINERs is done according to Eq.~\ref{eq: New blue SFG/AGN}.

\subsection{DEW diagnostic}

The DEW diagnostic has been proposed by \citet{2006MNRAS.371..972S}
and involves the DEW diagnostic diagram, showing the $D_{n}(4000)$
index vs. the maximum (in absolute value) of the equivalent widths
of $[\mbox{O}\textsc{ii}]$ and $\mathrm{[Ne\textsc{iii}]}$ emission
lines. We separate AGNs from SFGs using

\begin{equation}
D_{n}(4000)=-0.15x'+1.7,\label{eq: DEW Grazyna}
\end{equation}
with $x'=\mathrm{\log\left(max\left(EW[O\textsc{ii}],EW[Ne\textsc{iii}]\right)\right)}+1$,
where AGNs are above this line. This diagnostic is based the $D_{n}(4000)$
index being an indicator of the mean age of the stellar populations.
Thus, it is indeed useful to separate galaxies that are dominated
by older stars (AGNs) from galaxies dominated by younger stars (SFGs).
The DEW diagnostic also considers that the {[}Ne\noun{iii}{]} emission
line may be stronger than the $[\mbox{O}\textsc{ii}]$ emission line
in AGNs. Thus, it should be used as a good additional tracer for AGNs
in low signal-to-noise ratio surveys.

\subsection{Summary\label{sub:Summary}}

\begin{figure*}
\begin{centering}
\includegraphics[width=0.3\paperwidth]{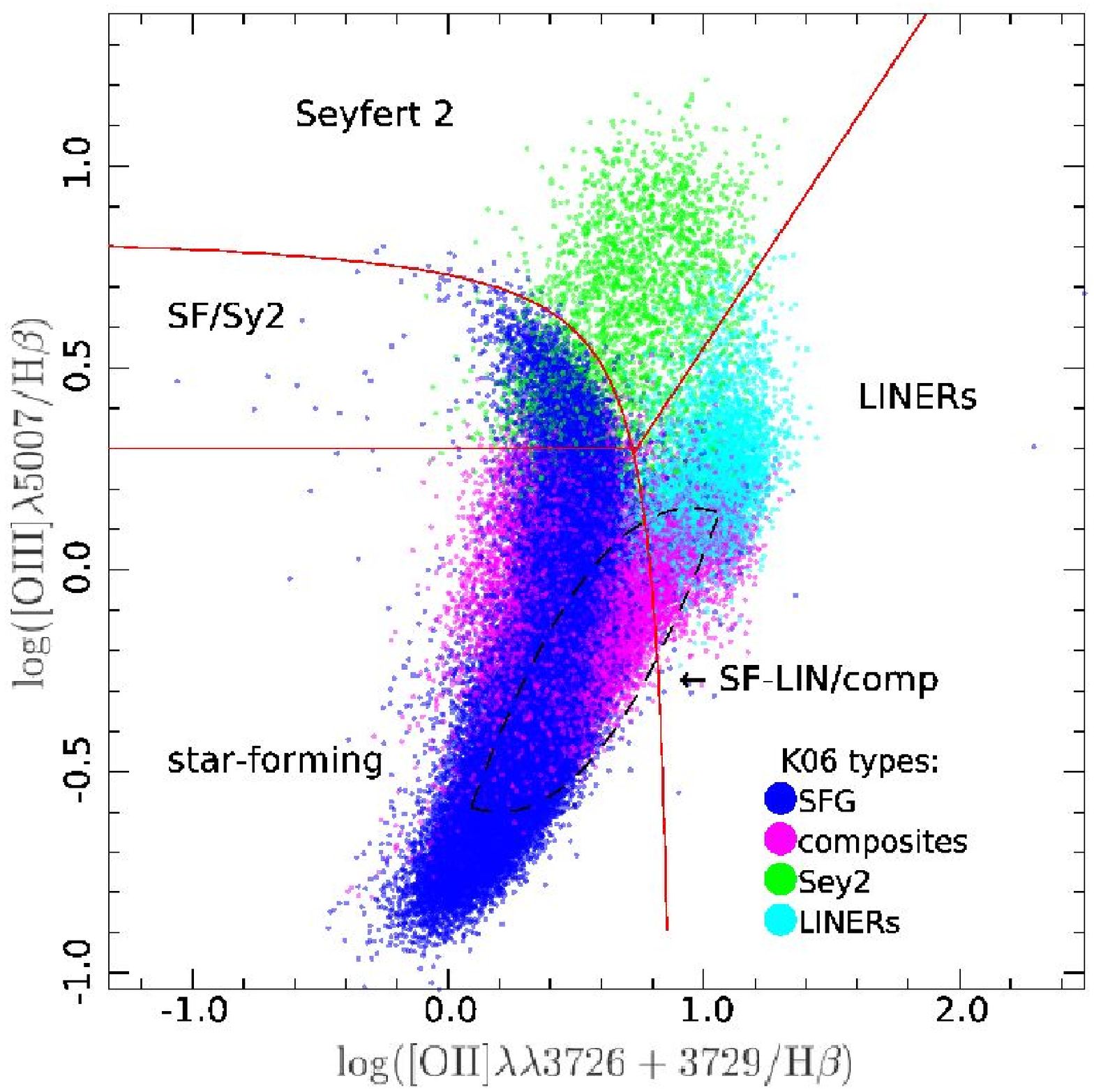} \includegraphics[width=0.3\paperwidth]{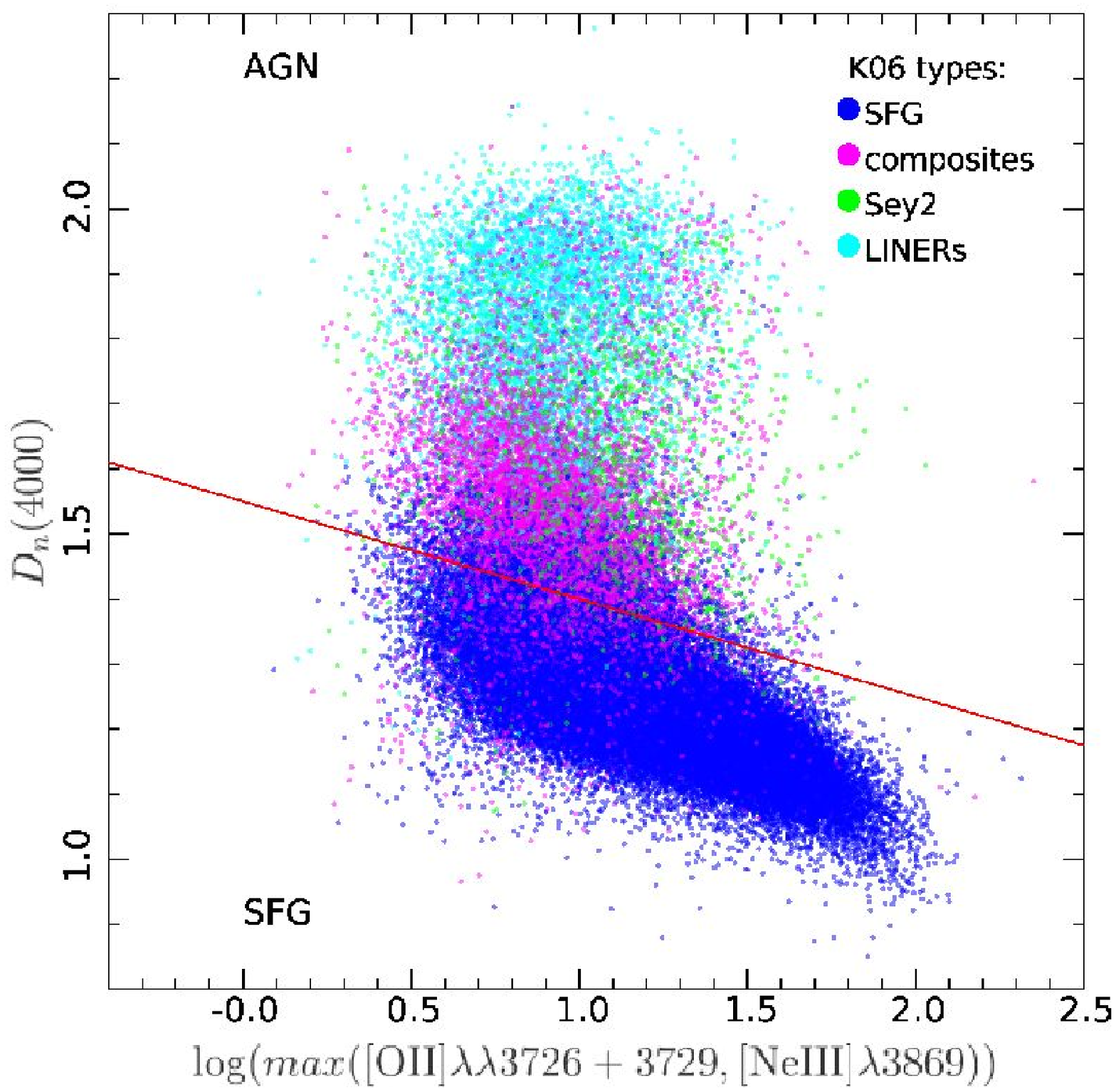}\\
\includegraphics[width=0.3\paperwidth]{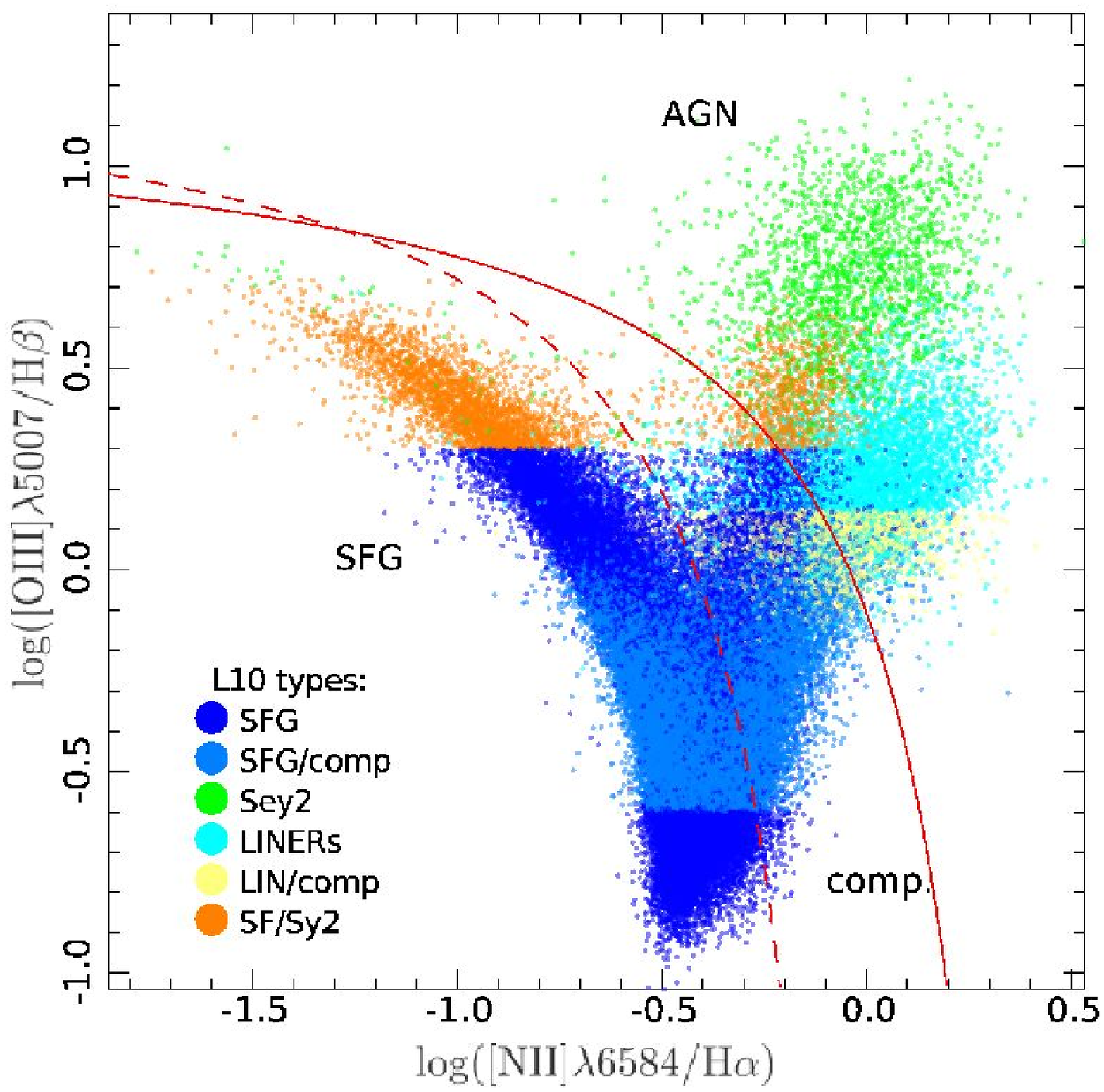} \includegraphics[width=0.3\paperwidth]{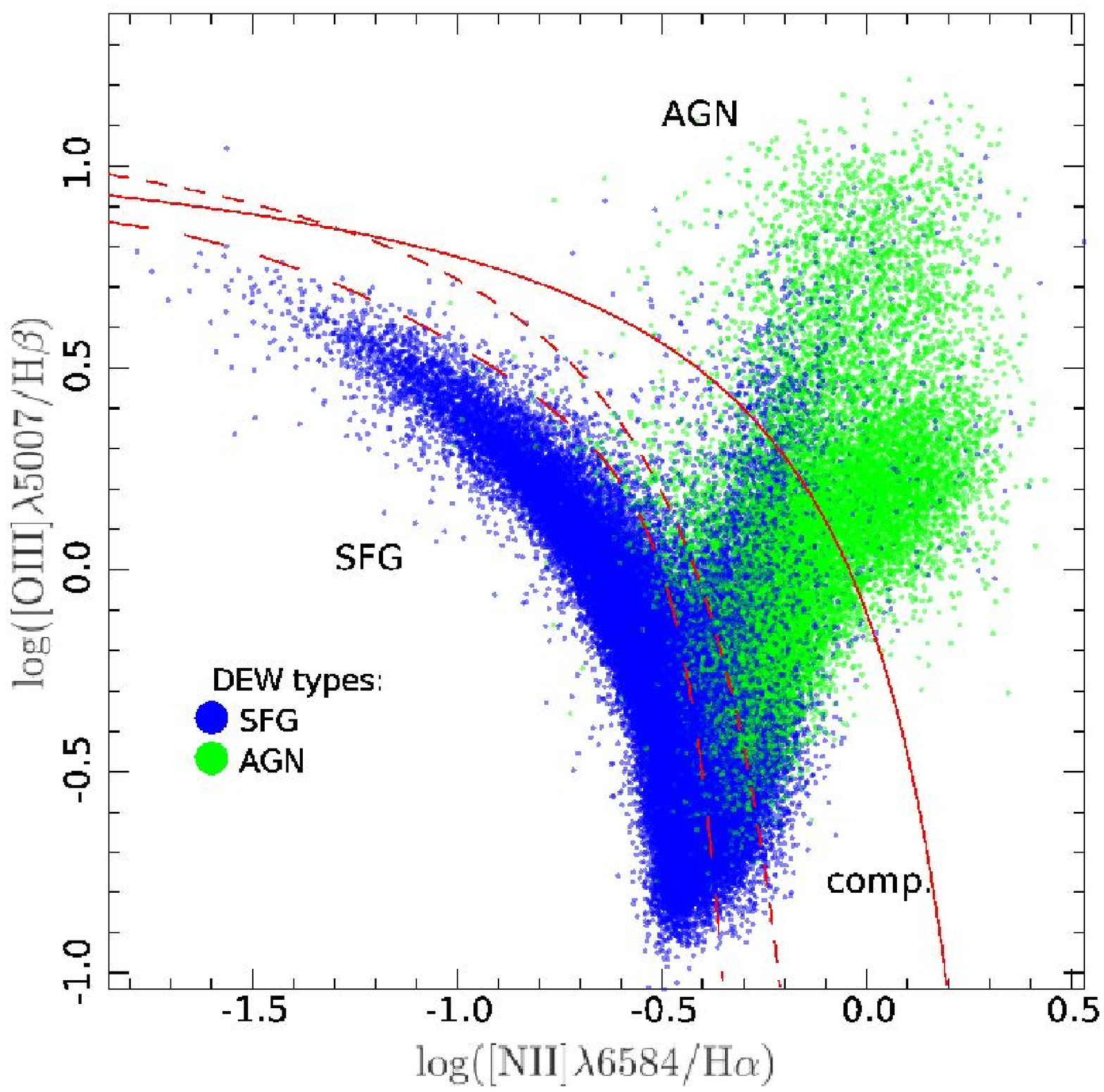}
\par\end{centering}

\caption{Summary of existing schemes for the classification of emission-line
galaxies at high redshift. \emph{Top}: results of the reference diagnostic
\citep{2006MNRAS.372..961K} are shown in the log$\left([\mathrm{O\textrm{\textsc{iii}}}]\lambda5007/\mathrm{H}\beta\right)$
vs. the $\log\left([\mathrm{O}\textrm{\textsc{ii}}]\lambda\lambda3726+3729/\mathrm{H}\beta\right)$
diagram (left) and in the $D_{n}(4000)$ vs. $\mathrm{max\left(EW[O\textsc{ii}],EW[Ne\textsc{iii}]\right)}$
diagram (right). The code is \emph{blue}: SFG; \emph{green}: Seyfert
2; \emph{cyan}: LINERs; \emph{magenta}: composites. \emph{Bottom}:
The results of the high-redshift classifications are shown in the
standard log$\left([\mathrm{O\textrm{\textsc{iii}}}]\lambda5007/\mathrm{H}\beta\right)$
vs. log$\left([\mathrm{N}\textsc{ii}]\lambda6583/\mathrm{H}\alpha\right)$
diagram. \emph{Left}: the L10 diagnostic \citep{2009lama}. \emph{Right}:
the DEW diagnostic \citep{2006MNRAS.371..972S}. Same color code as
above except \emph{green}: stands for all AGNs in the bottom-right
panel; \emph{light blue}: SFG/comp; \emph{yellow}: LIN/comp; \emph{orange}:
SF/Sy2.In bottom-right panel, the long dashed line is the boundary
between SFG and AGN used by \citet{2006MNRAS.371..972S}.}

\label{Flo:summary}
\end{figure*}

Figure~\ref{Flo:summary} shows how the different types of galaxies
(according to K06) appear in the high-redshift diagrams (top panels)
and how the high redshift classifications appear back in one of the
K06 diagnostic diagrams (bottom panels). In all panels, SFG are plotted
in blue, Seyfert 2 in green (except in the bottom-right panel where
green points stand for all types of AGNs), LINERs in cyan, and composites
in magenta. The L10 diagnostic (left panels) implies several regions
where different types of galaxies get mixed. Seyfert 2 region and
LINERs region are now quite well defined, but we see composites falling
in the SFGs and LINERs regions. Most of the composites fall in the
region of the L10 diagnostic called SF-LIN/comp(marked by the dashed
contour corresponding to Eqs.~\ref{eq: New blue composite 1} and~\ref{eq: New blue composite 2}).
SFGs and Seyfert 2 are now separated quite well, but still there is
a small region of the L10 diagnostic, called SF/Sy2, where they get
mixed. In the bottom-left panel, it seems that most of the SF/Sy2
galaxies belong to the K06 SFG region, and that a large number of
SFG/comp galaxies belong to K06 SFG region. LIN/comp galaxies seem
to appear half/half in the K06 composites and LINERs regions.

We now the compare K06 and DEW classifications (right panels). We
see that all K06 LINERs are correctly classified as AGNs in the DEW
diagnostic. Most of K06 Seyfert 2 galaxies lie in the DEW AGN region
as well, so that is quite satisfying. However composites are shared
in DEW SFG and AGN regions, which confirms that composites are sort
of hybrids between AGNs and SFGs, also in terms of stellar populations.
Thus they obviously cannot be isolated in the DEW diagnostic. We emphasize
that the definition of SFG and AGN galaxies used in \citet{2006MNRAS.371..972S}
is slightly different from the K06 scheme: it is based on the long-dashed
curve shown in the bottom-right panel of Fig.~\ref{Flo:summary}
(Eq.~11 of their paper). The DEW diagnostic is designed to exclude
only pure SFGs without any AGN contribution, while according to \citet{2006MNRAS.371..972S}
galaxies classified as SFG by K06 (and by us) would allow up to 3\%
AGN contribution. Composites would allow up to 20\% AGN contribution.

Indeed in the bottom-right panel, a non-negligible number of DEW AGNs
actually belong to the K06 SFG or composite regions.However, we also
note conversely that a non-negligible number of DEW SFGs contaminate
the K06 composite and AGN regions. The DEW diagnostic actually fails
to completely exclude all pure SFGs.

\section{Limits of the L10 and DEW classifications\label{sec: Limits of blue classifications}}

\subsection{Success and contamination charts\label{sub:Success-and-contamination}}

\begin{table}
\caption{Success chart of the L10 diagnostic, where numbers are the probability
that a given reference K06 type goes in a given L10 type. }

\begin{lyxlist}{00.00.0000}
\item [{%
\begin{tabular}{ccccc}
\hline 
\hline
 & \multicolumn{4}{c}{reference K06 classification}\tabularnewline
\hline 
L10 classification & SFG & Composites & Seyfert 2 & LINERs\tabularnewline
\hline 
\emph{total} & \emph{100} & \emph{100} & \emph{100} & \emph{100}\tabularnewline
Seyfert 2 & 0.08 & 0.19 & 59.27 & 3.81\tabularnewline
SFG/Sy2 & 3.94 & 1.13 & 26.25 & 0.12\tabularnewline
SFG & 40.24 & 27.79 & 5.90 & 2.44\tabularnewline
SFG/comp & 55.47 & 55.15 & 0.14 & 2.93\tabularnewline
\emph{total SFG}\emph{\small $^{*}$} & \emph{99.65} & \emph{84.07} & \emph{32.28} & \emph{5.50}\tabularnewline
LINERs & 0.17 & 6.94 & 8.00 & 73.51\tabularnewline
LIN/comp & 0.11 & 8.81 & 0.44 & 17.18\tabularnewline
\emph{total LIN}\emph{\small $^{**}$} & \emph{0.27} & \emph{15.74} & \emph{8.44} & \emph{90.70}\tabularnewline
\hline 
\end{tabular}}]~
\end{lyxlist}
{\small $^{*}$ union of SFG/Sy2, SFG, and SFG/comp regions.}{\small \par}

{\small $^{**}$ union of the LINERs and LIN/comp regions.}{\small \par}

\label{Flo: Succes chart New blue}
\end{table}
\begin{table}
\caption{Contamination chart for the L10 diagnostic, where numbers are the
probability that a given L10 type actually is any of the reference
K06 types. }

\begin{lyxlist}{00.00.0000}
\item [{%
\begin{tabular}{cccccc}
\hline 
\hline
 & \multicolumn{5}{c}{reference K06 classification}\tabularnewline
\hline 
L10 classification & \emph{total} & SFG & Comp. & Seyfert 2 & LINERs\tabularnewline
\hline 
Seyfert 2 & \emph{100} & 2.73 & 1.29 & 86.71 & 9.28\tabularnewline
SFG/Sy2 & \emph{100} & 74.05 & 4.30 & 21.48 & 0.17\tabularnewline
SFG & \emph{100} & 86.90 & 12.17 & 0.55 & 0.38\tabularnewline
SFG/comp & \emph{100} & 82.96 & 16.72 & 0.01 & 0.32\tabularnewline
\emph{total SFG}\emph{\small $^{*}$} & \emph{100} & \emph{84.10} & \emph{14.38} & \emph{1.19} & \emph{0.34}\tabularnewline
LINERs & \emph{100} & 2.28 & 19.40 & 4.80 & 73.51\tabularnewline
LIN/comp & \emph{100} & 3.37 & 56.57 & 0.61 & 39.46\tabularnewline
\emph{total LIN}\emph{\small $^{**}$} & \emph{100} & \emph{2.61} & \emph{30.68} & \emph{3.53} & \emph{63.18}\tabularnewline
\hline 
\end{tabular}}]~
\end{lyxlist}
{\small $^{*}$ union of SFG/Sy2, SFG, and SFG/comp regions.}{\small \par}

{\small $^{**}$ union of the LINERs and LIN/comp regions.}{\small \par}

\label{Flo: Contamination chart New blue}
\end{table}

The success chart consists in classifying galaxies from our sample
according to the reference, then associating a probability for each
type of galaxy (AGN, composite, or SFG) to be classified correctly
in the new diagnostic. The contamination chart is based on the same
principle as the success chart, except this time we classify galaxies
according to the new diagnostic, and then we calculate the probability
that the galaxies classified as one type are actually of that same
type according to the reference. Table~\ref{Flo: Succes chart New blue}
shows the success chart of the L10 diagnostic. It reveals a relatively
satisfying spread of composite galaxies and AGNs inside the different
types defined. Table~\ref{Flo: Contamination chart New blue} shows
the associated contamination chart. We notice quite good efficiency,
i.e. low contamination by other types, in the L10 SFG, Seyfert 2,
and LINER regions.

If we take a look at AGNs, we notice that almost 60\% of K06 Seyfert
2 galaxies are successfully classified as L10 Seyfert 2. Moreover
26\% belong to the L10 SFG/Sy2 region, which would give us a total
of more than 85\% of K06 Seyfert 2 galaxies being classified as Seyfert2
with the L10 diagnostic. However, the contamination chart shows that
the L10 SFG/Sy2 region is actually made up of only 21\% K06 Seyfert
2, which means it cannot be used to reliably look for additional Seyfert
2 galaxies in high-redshift samples. The Seyfert 2 region itself shows
a very low contamination (13\%) by other types.

Most K06 LINERs (74\%) are also successfully classified as L10 LINERs,
and 17\% belong to the L10 LIN/comp region. That gives a global success
rate of 91\% for LINERs. As already stated in Paper~I, these results
are much better than results produced by the former L04 diagnostic;
however, the contamination by composites in the LIN/comp region is
not negligible. Only 40\% of the L10 LIN/comp objects actually are
K06 LINERs, and 57\% are K06 composites. That gives a global 37\%
contamination in the union of the L10 LINERs and LIN/comp regions.

Finally we confirm the conclusion of Paper~I from these two tables,
which is that the L10 diagnostic is very efficient for SFGs. If we
consider the union of the L10 SFG, SFG/Sy2, and SFG/comp regions,
the success rate is 99.7\% and the contamination by other types only
16\%. This low contamination by composite galaxies and AGNs, in particular
in the SFG/Sy2 and SFG/comp regions, has been shown to not critically
bias SFG studies such as metallicity. \citet{2009A&A...495...53L},
for instance, performed such tests using the L04 diagnostic, i.e.
with an even more contamination by AGNs in the SFG region.

\begin{table}
\caption{Success chart for DEW diagnostic}

\begin{lyxlist}{00.00.0000}
\item [{%
\begin{tabular}{ccccc}
\hline 
\hline
 & \multicolumn{4}{c}{reference K06}\tabularnewline
\hline 
DEW & SFG & Composites & Seyfert 2 & LINERs\tabularnewline
\hline 
\emph{total} & \emph{100} & \emph{100} & \emph{100} & \emph{100}\tabularnewline
SFG & 94.95 & 38.08 & 20.24 & 2.16\tabularnewline
AGN & 5.05 & 61.92 & 79.76 & 97.84\tabularnewline
\hline 
\end{tabular}}]~
\end{lyxlist}
\label{Flo: Success chart DEW}
\end{table}

\begin{table}
\caption{Contamination chart for DEW diagnostic}

\begin{lyxlist}{00.00.0000}
\item [{%
\begin{tabular}{cccccc}
\hline 
\hline
 & \multicolumn{5}{c}{reference K06}\tabularnewline
\hline 
DEW & \emph{total} & SFG & Composites & Seyfert 2 & LINERs\tabularnewline
\hline 
SFG & \emph{100} & 91.56 & 7.44 & 0.85 & 0.15\tabularnewline
AGN & \emph{100} & 17.94 & 44.57 & 12.32 & 25.17\tabularnewline
\hline 
\end{tabular}}]~
\end{lyxlist}
\label{Flo: Contamination chart DEW}
\end{table}

Tables~\ref{Flo: Success chart DEW} and~\ref{Flo: Contamination chart DEW}
show the success and contamination charts of the DEW diagnostic. Again,
the results are very good for SFGs with a 95\% success rate, and only
an 8\% contamination, less than for L10. Still, this better contamination
chart for SFGs does not drastically reflect a worse success rate for
AGNs. The success rate is indeed 80\% for Seyfert 2 and 98\% for LINERs.
The DEW has in fact greater ability to separate SFGs from AGNs than
standard diagnostic diagrams as in K06 or L10 classifications. However,
the main limitations of the DEW diagnostic clearly appear from the
contamination chart regarding DEW AGNs. The DEW AGN region is actually
made up of only 37\% K06 AGNs. There is indeed a high contamination
by 18\% K06 SFGs, much higher than with the L10 diagnostic (less than
3\% in the Seyfert 2 and LINERs regions). Moreover, the DEW AGN region
is contaminated by 45\% K06 composites, to be compared to 30\% for
the L10 LINERs, and only 1\% for the L10 Seyfert 2. As one can see
in Fig.~\ref{Flo:summary}, composites get completely confused with
Seyfert 2 and LINERs in the DEW diagnostic, while they are rather
confused with SFGs in the L10 diagnostic. We do note that this contamination
is explained mainly by the fact that \citet{2006MNRAS.371..972S}
use a different definition of SFGs and AGN galaxies (as discussed
in Sect.~\ref{sub:Summary} above). Indeed, in DEW diagnostic's philosophy,
it should not be considered as a {}``contamination'' but as a {}``contribution''
of an AGN to star-forming galaxies. Finally, the DEW diagnostic does
not allow any distinction between Seyfert 2 and LINERs.

Regarding the classification of SFGs, we conclude that one should
use the L10 diagnostic for its very high success rate and DEW diagnostic
for its lower contamination. About AGNs, both advantages of the L10
and DEW diagnostic diagrams may be put together to provide a better
diagnostic, which is the goal of the present paper. We emphasize that
we did not use the DEW diagnostic itself for our own classification.
We only used the DEW diagram to derive a new diagnostic where it is
needed (see Sect.~\ref{sub: New blue classification} below).

\subsection{AGN counts\label{sub: New blue classification}}

\begin{figure*}
\begin{centering}
\includegraphics[width=0.2\paperwidth]{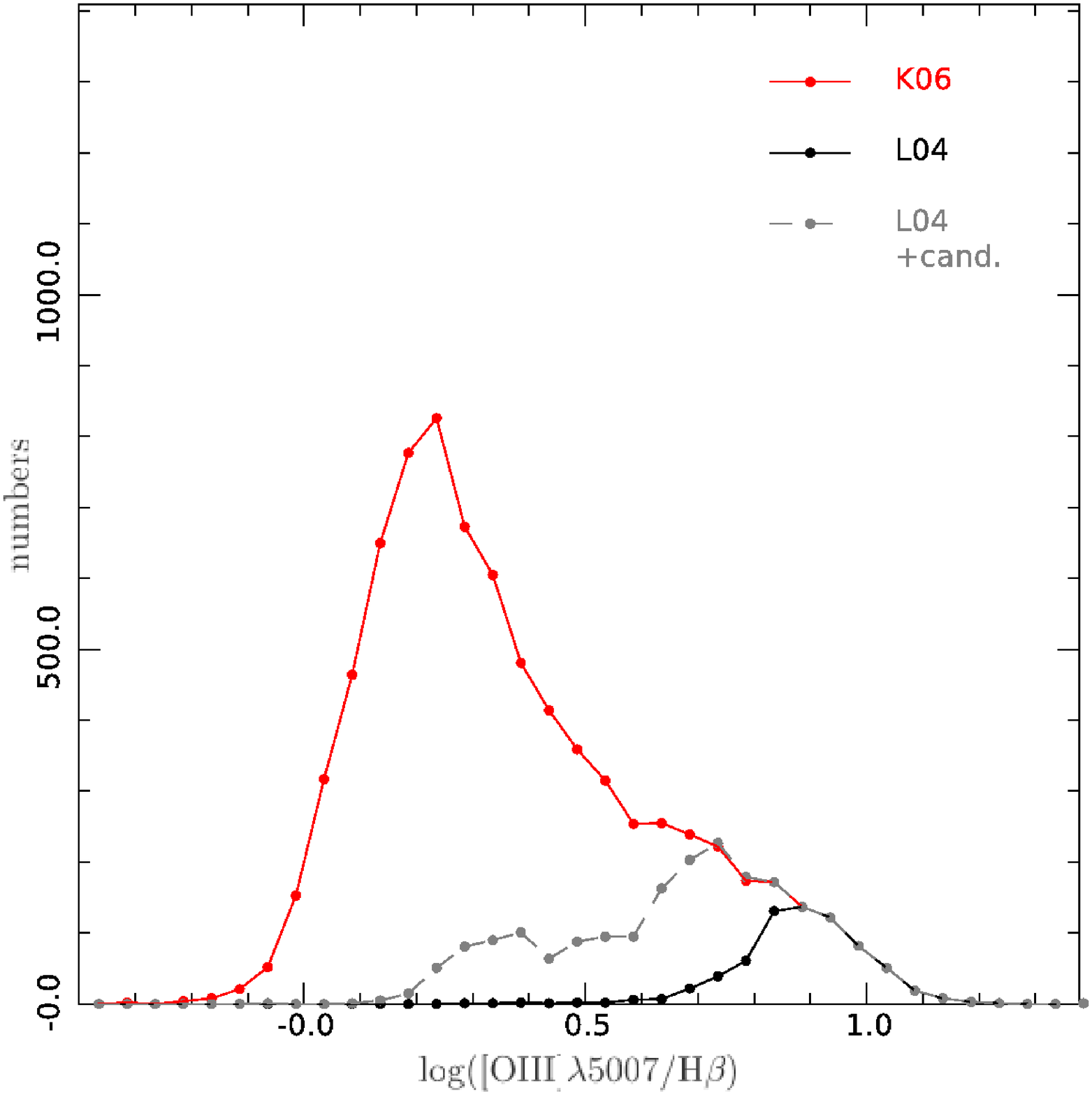} \includegraphics[width=0.2\paperwidth]{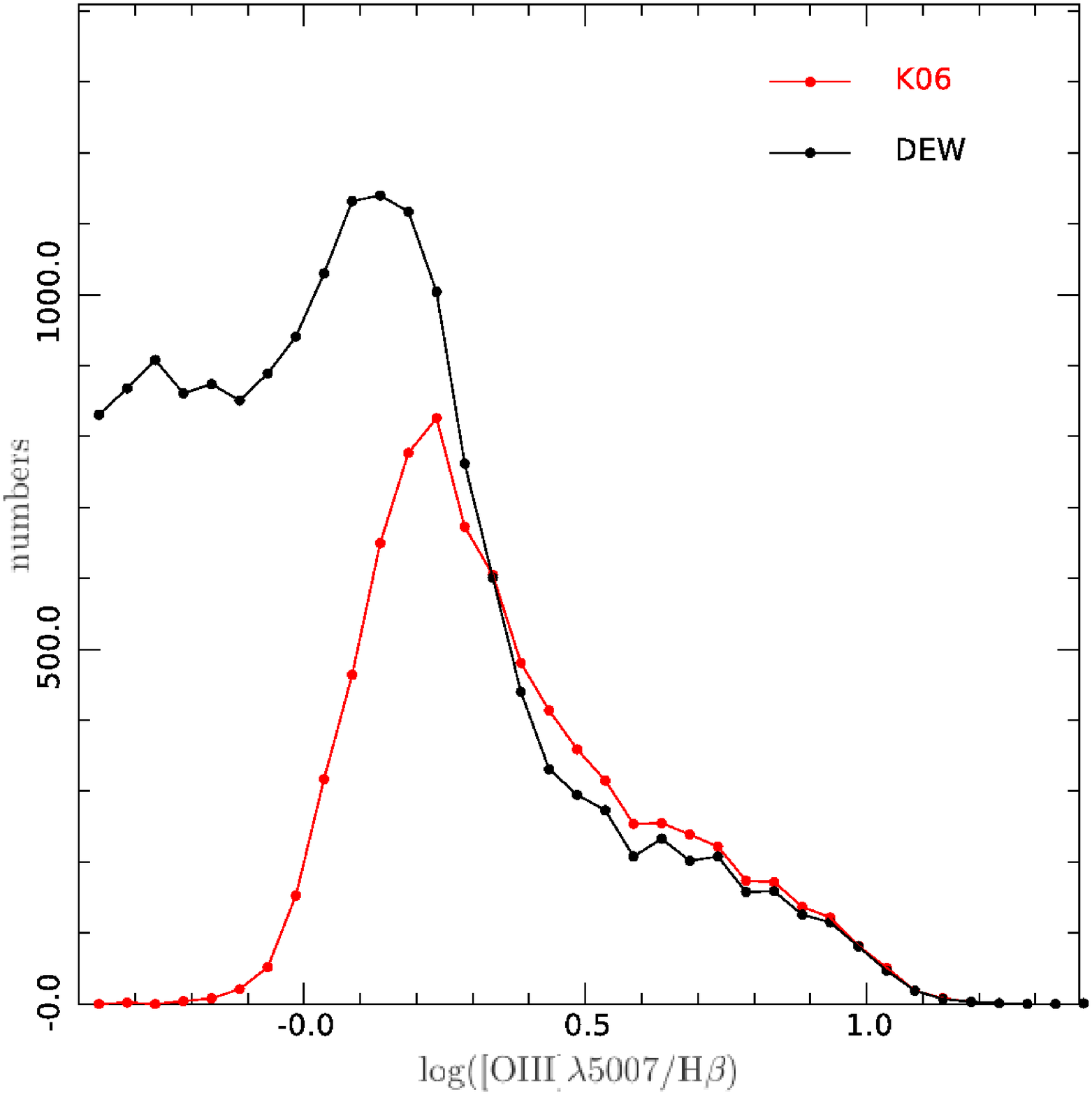}\includegraphics[width=0.2\paperwidth]{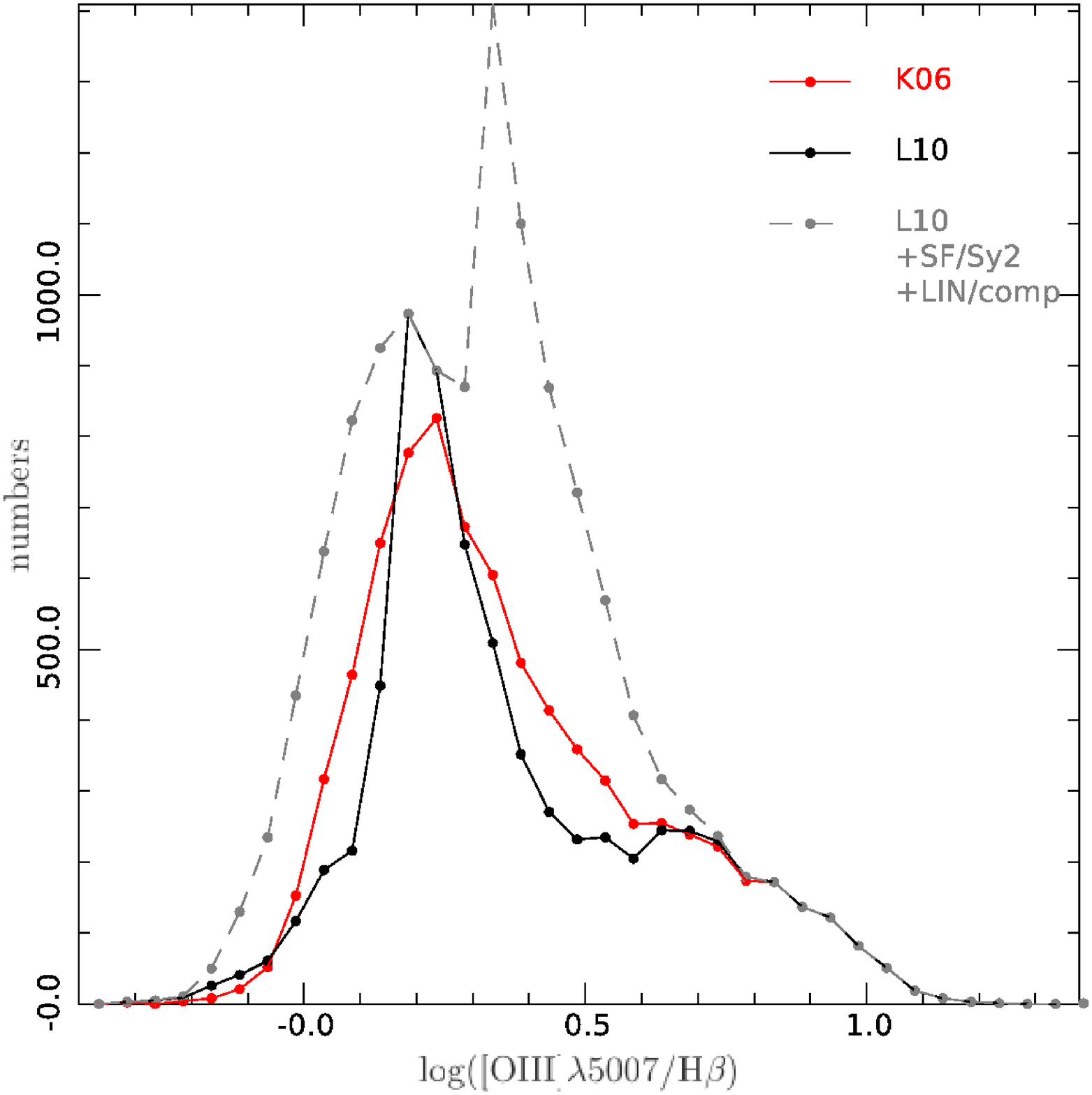}\includegraphics[width=0.2\paperwidth]{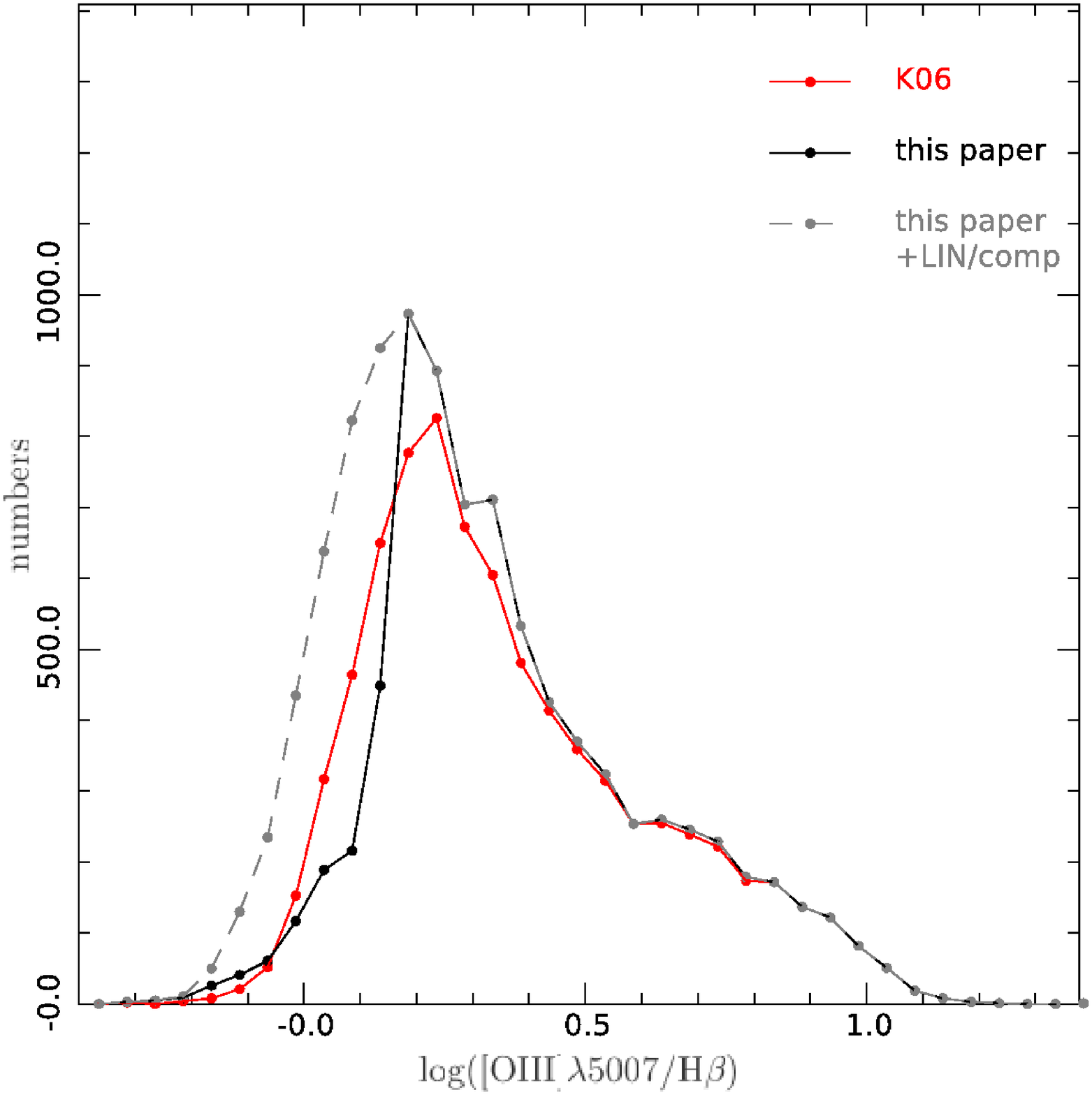}
\par\end{centering}

\begin{centering}
\includegraphics[width=0.2\paperwidth]{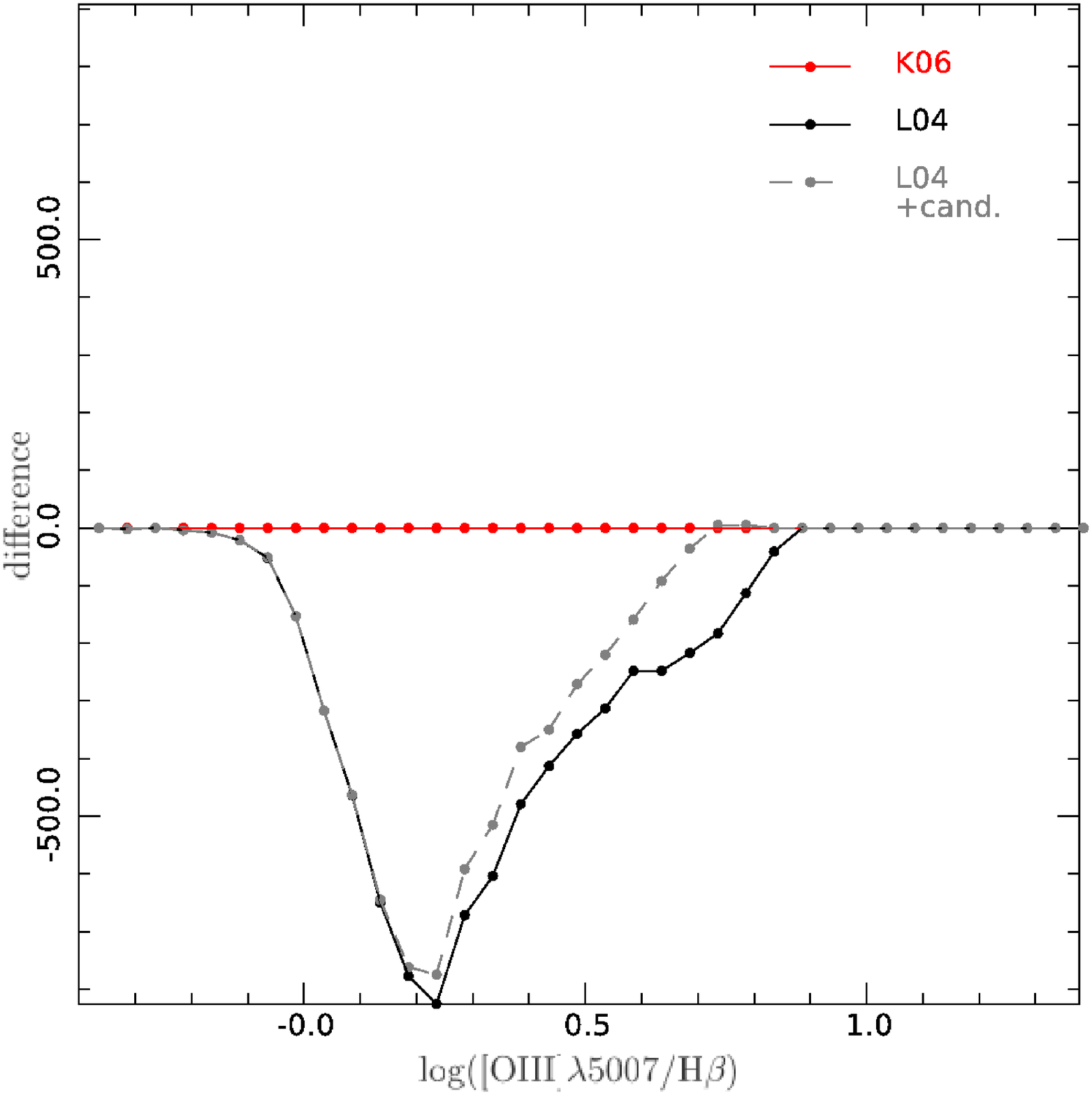} \includegraphics[width=0.2\paperwidth]{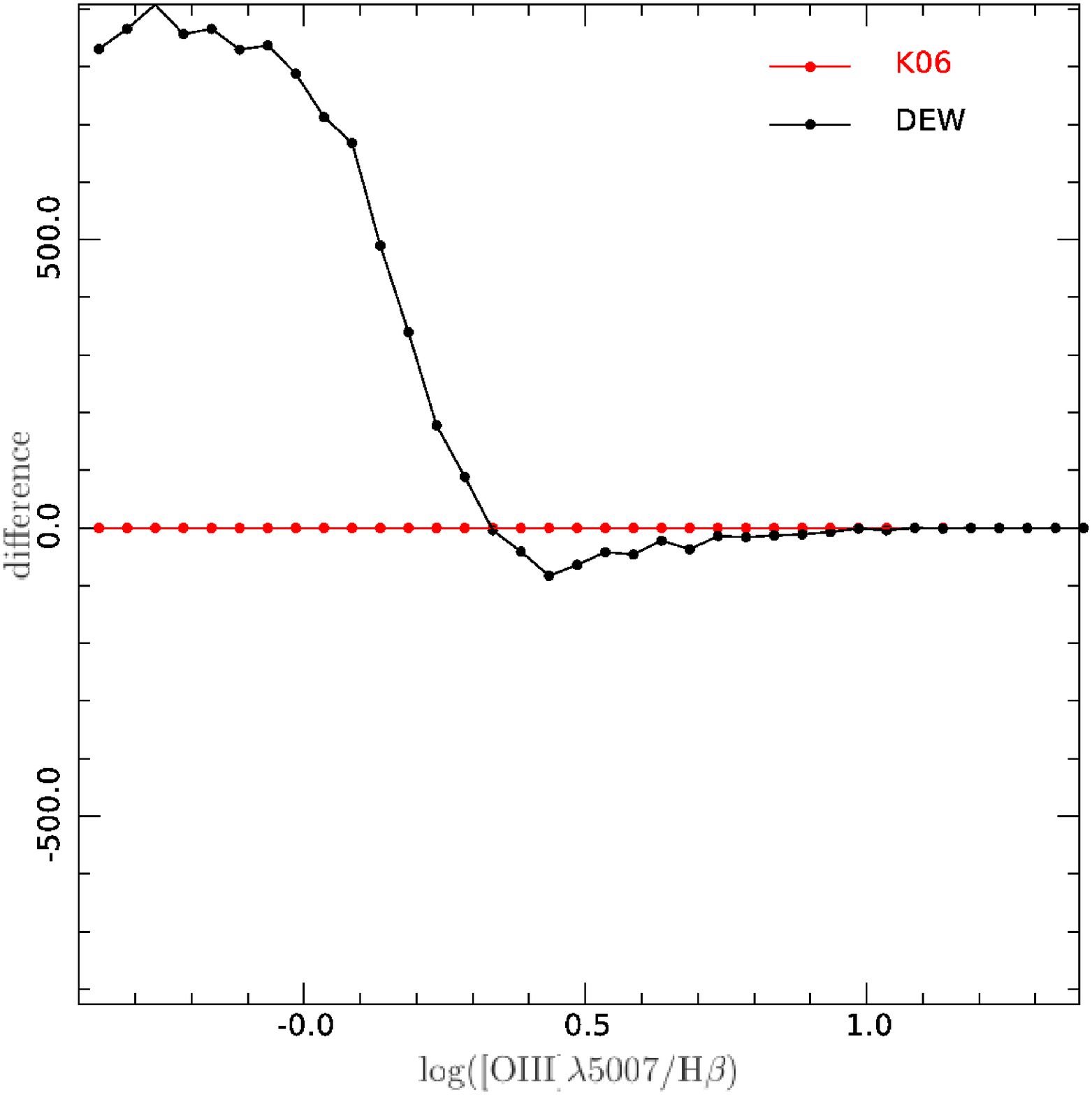}\includegraphics[width=0.2\paperwidth]{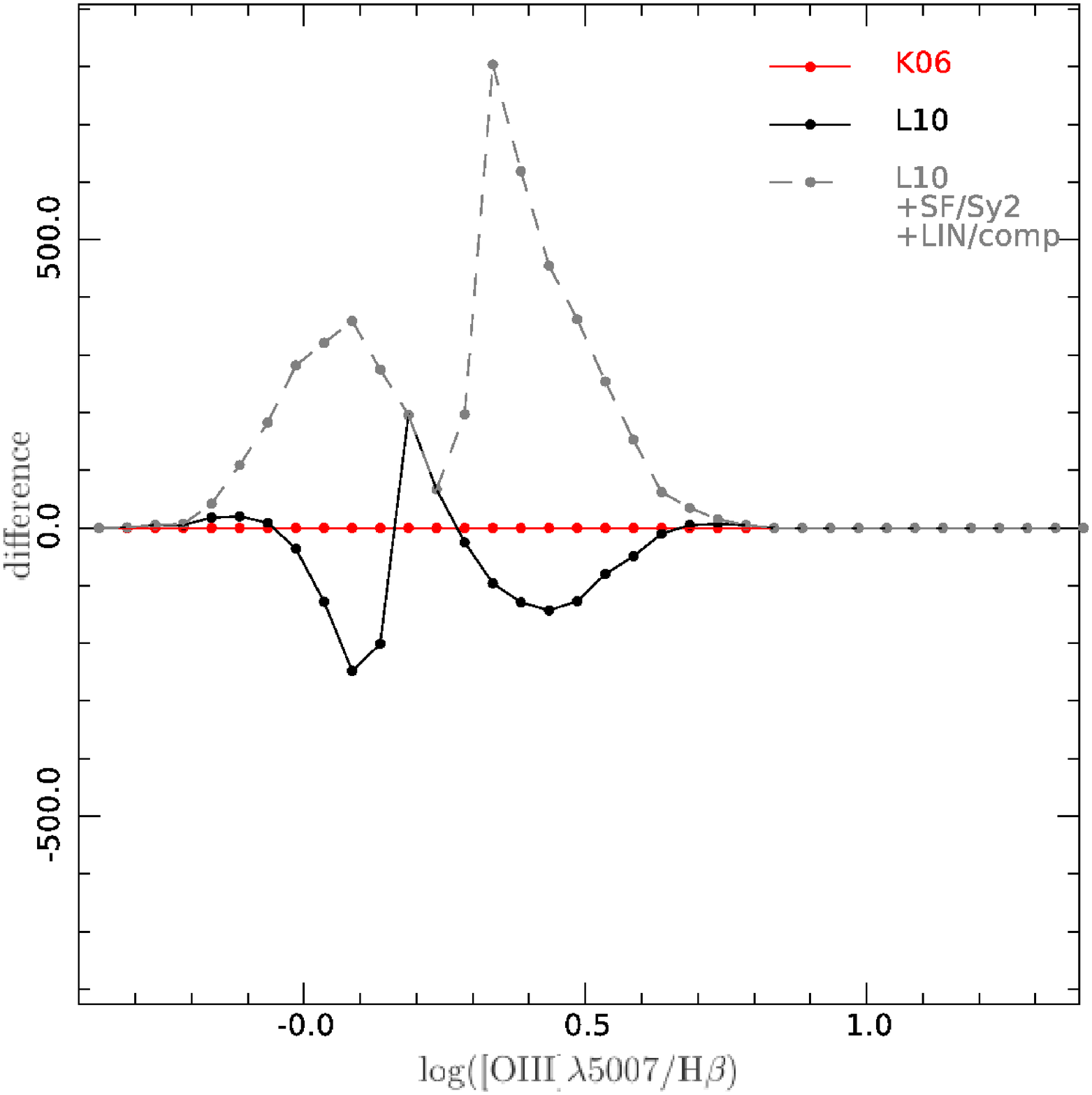}\includegraphics[width=0.2\paperwidth]{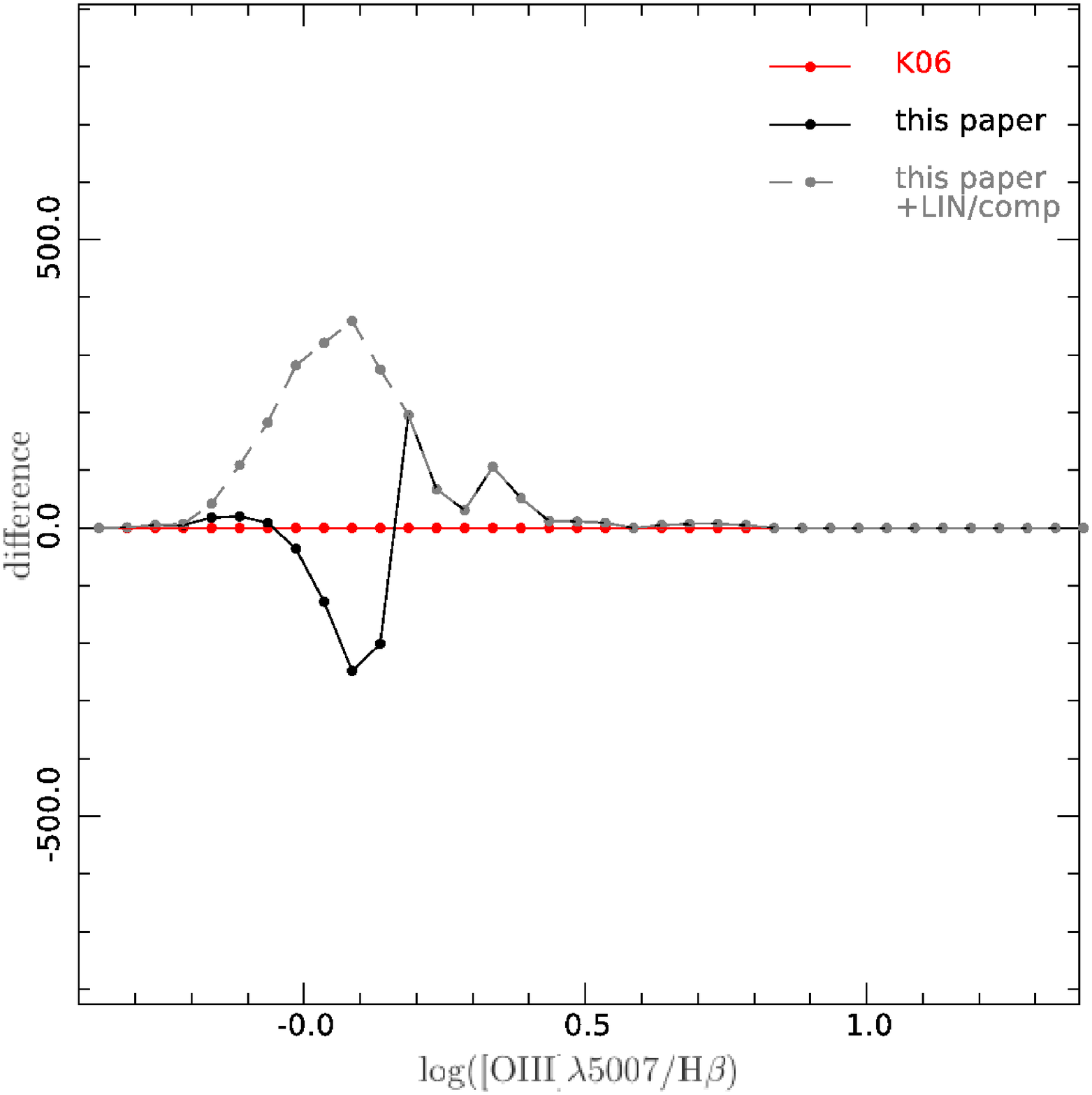}
\par\end{centering}

\caption{Comparison between the number of AGNs (Seyfert 2 and LINERs) counted
with different calibrations at high redshift, as a function of log$\left([\mathrm{O\textrm{\textsc{iii}}}]\lambda5007/\mathrm{H}\beta\right)$.
From left to right, the calibrations used are L04, DEW, L10 (paper~I),
and the present paper. Top panels show absolute counts. Bottom panels
show difference counts. In each panel, the reference counts established
with the K06 diagnostic are shown in red, and the counts obtained
with the high-redshift diagnostic are shown in black. The gray dashed
lines show the results when including {}``candidate'' regions, i.e.
for L04, L10, and the present paper's classifications: candidate AGNs
region; SFG/Sy2 and LIN/comp regions; LIN/comp region.}

\label{Flo: AGN count newblue}
\end{figure*}

In order to explore the limitations of L10 and DEW classifications
of AGNs better, we now count the number of AGNs (Seyfert 2 and LINERs)
as a function of the ionization state, roughly given by the $\log\left([\mathrm{O}\textsc{iii}]/\mathrm{H}\beta\right)$
emission-line ratio. To achieve this test, we divide the K06 or L10
diagnostic diagrams in equal horizontal slices, and then in each slice
we count the number of AGNs. Figure~\ref{Flo: AGN count newblue}
shows the absolute and difference counts (relative to K06) obtained
with, from left to right, the following classifications: L04, DEW,
L10, and the present paper's diagnostic (see Sect.~\ref{sec:Spectral-classification}
below). In each panel, the results are compared to the actual count
of AGNs according to the reference K06 diagnostic.

We confirm, as stated in Paper~I, that the L04 diagnostic tends to
underestimate the amount of AGNs, even when including L04 candidate
AGNs, and that a very high number of AGNs (mainly LINERs) are lost
in this diagnostic. However, we can put this effect into context thanks
to Fig.~\ref{Flo: AGN count newblue}, as we see that it only becomes
significant for $\log\left([\mathrm{O}\textsc{iii}]/\mathrm{H}\beta\right)\lesssim0.9$,
or $\log\left([\mathrm{O}\textsc{iii}]/\mathrm{H}\beta\right)\lesssim0.7$
if we include candidate AGNs. For AGNs with a high ionization state,
the L04 diagnostic indeed gives perfect results.

Figure~\ref{Flo: AGN count newblue} also shows that DEW and L10
classifications are doing quite well by following the K06's curve
almost exactly for $\log\left([\mathrm{O}\textsc{iii}]/\mathrm{H}\beta\right)\gtrsim0.25$.
In both cases, we notice an underestimate of the number of AGNs for
$0.25\lesssim\log\left([\mathrm{O}\textsc{iii}]/\mathrm{H}\beta\right)\lesssim0.7$,
where this effect is more significant for the L10 diagnostic than
for the DEW diagnostic. Nevertheless, the DEW diagnostic clearly overestimates
the number of AGNs in low ionization states, i.e. mainly LINERs ($\log\left([\mathrm{O}\textsc{iii}]/\mathrm{H}\beta\right)\lesssim0.25$).
In this region, the L10 diagnostic, in contrast, satisfyingly follows
the K06's curve, with a small underestimate. 

Unfortunately, including the SFG/Sy2 and LIN/comp regions does not
help. It makes a peak of galaxies at $\log\left([\mathrm{O}\textsc{iii}]/\mathrm{H}\beta\right)\simeq0.4$
appear that does not fit the reference profile, while in a lower ionization
state the number of AGNs is now clearly overestimated. Those two effects
are from the high contamination of the L10 SFG/Sy2 region by K06 SFGs
and of the L10 LIN/comp region by K06 composites.

\section{The supplementary M11 diagnostic\label{sec:Spectral-classification}}

\begin{figure*}
\begin{centering}
\includegraphics[width=0.29\paperwidth]{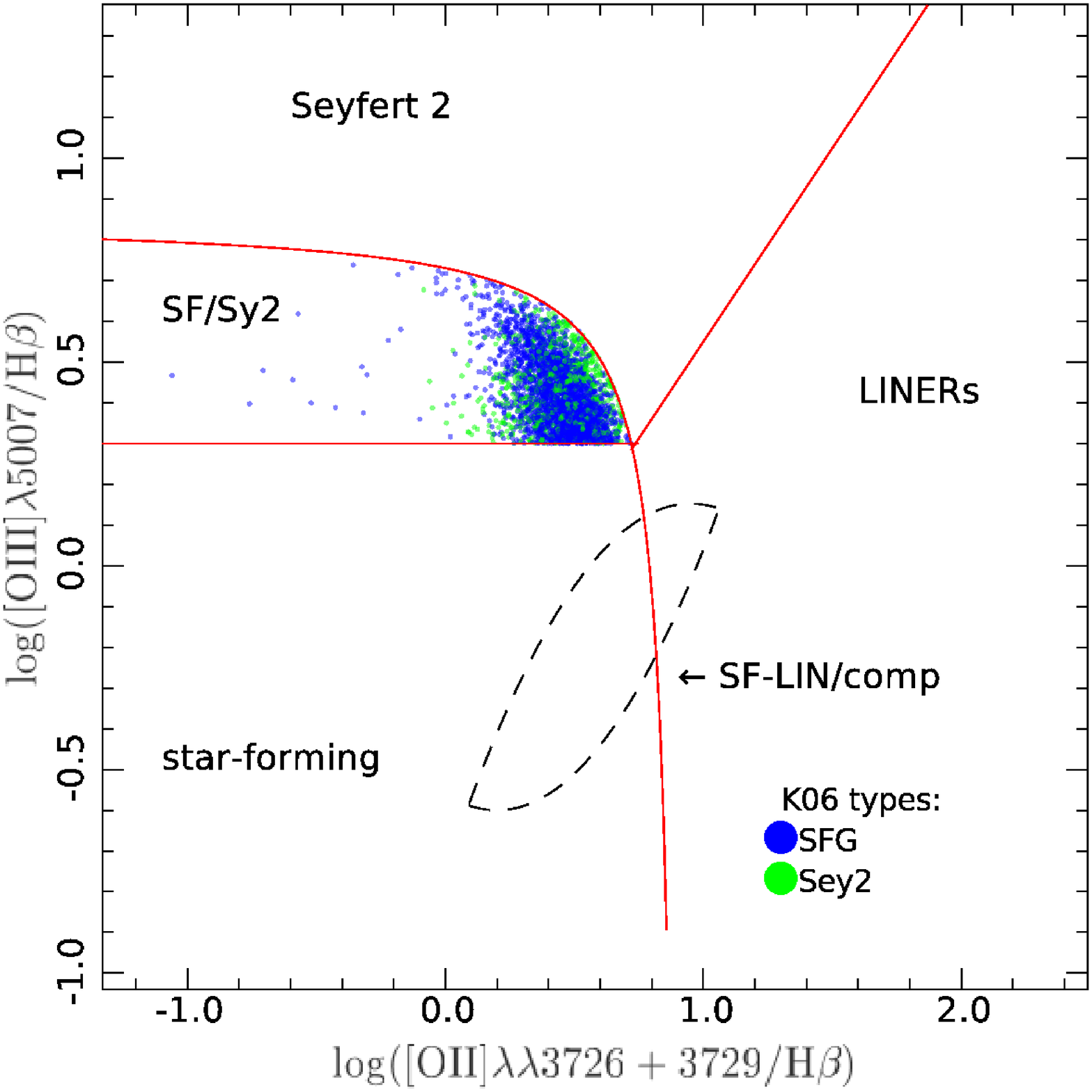}\includegraphics[width=0.29\paperwidth]{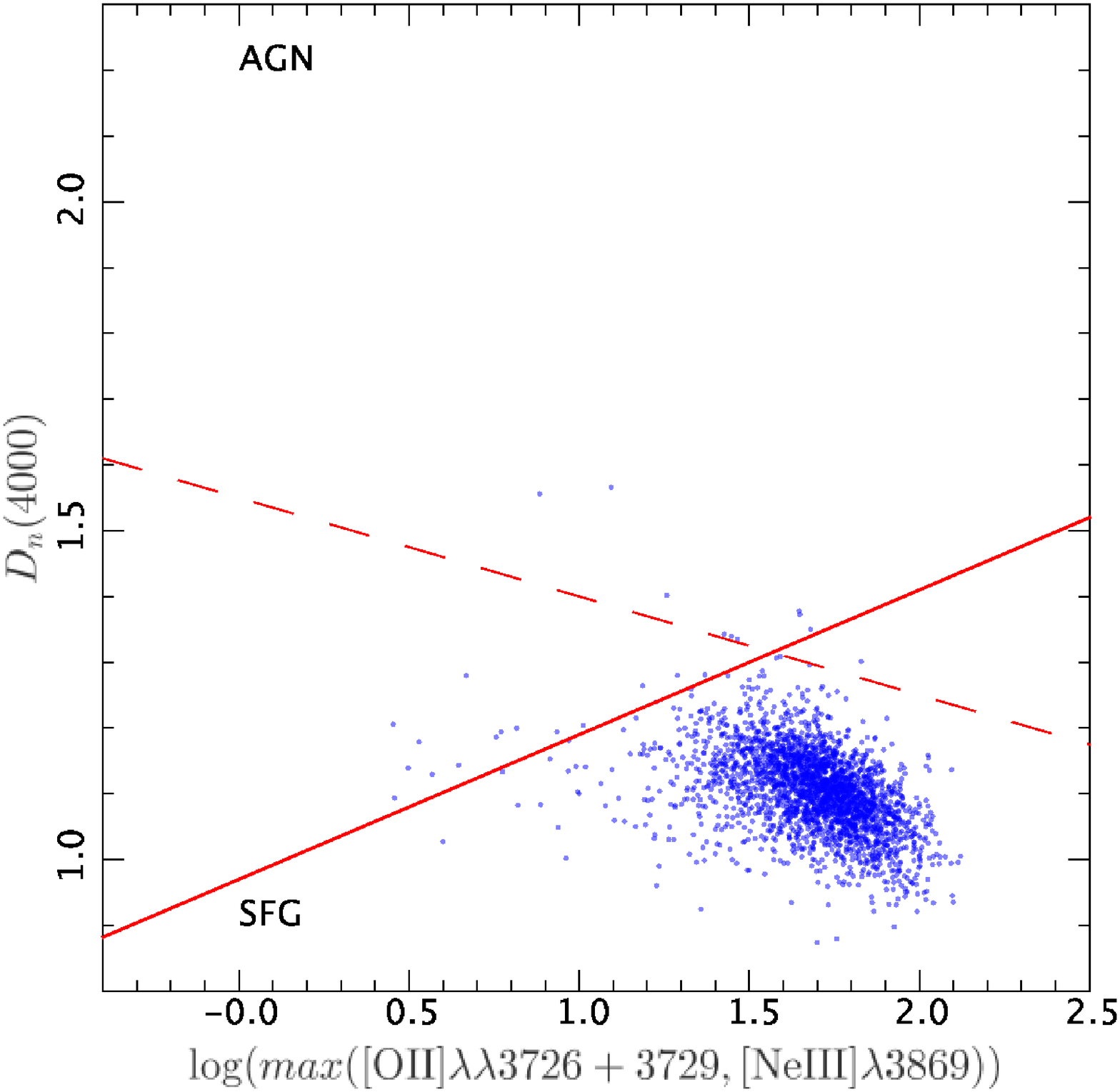}\includegraphics[width=0.29\paperwidth]{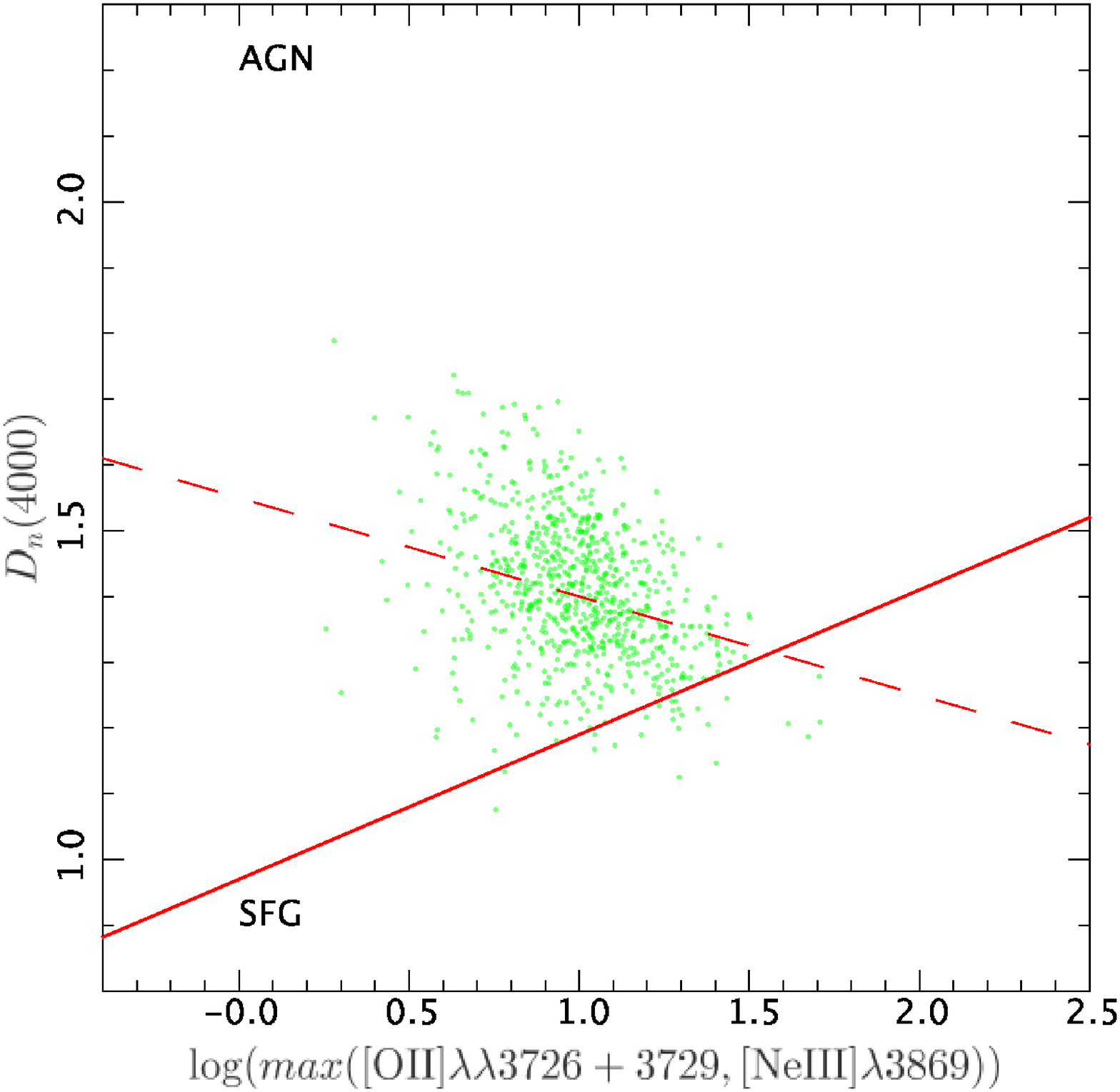}
\par\end{centering}

\begin{centering}
\includegraphics[width=0.29\paperwidth]{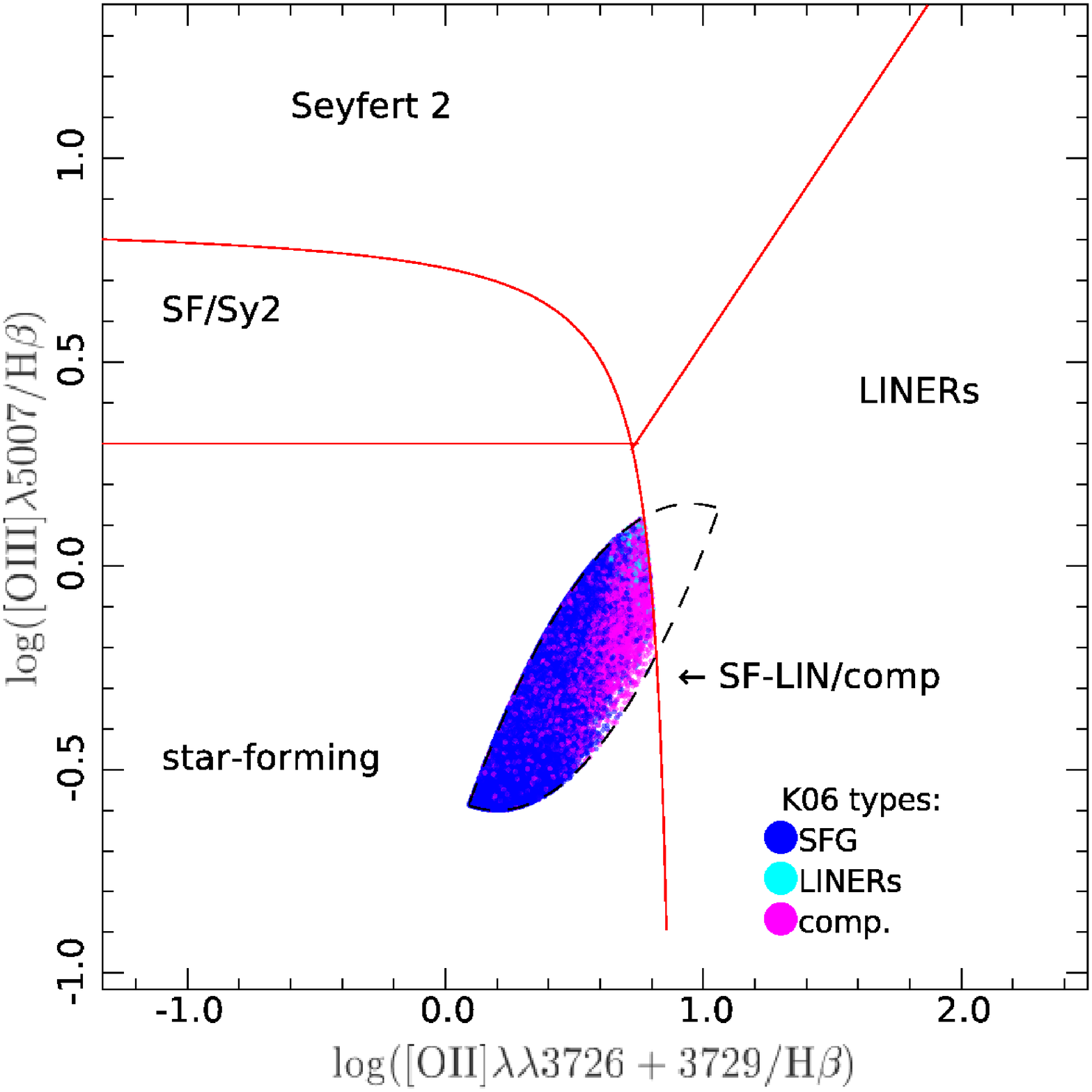}\includegraphics[width=0.29\paperwidth]{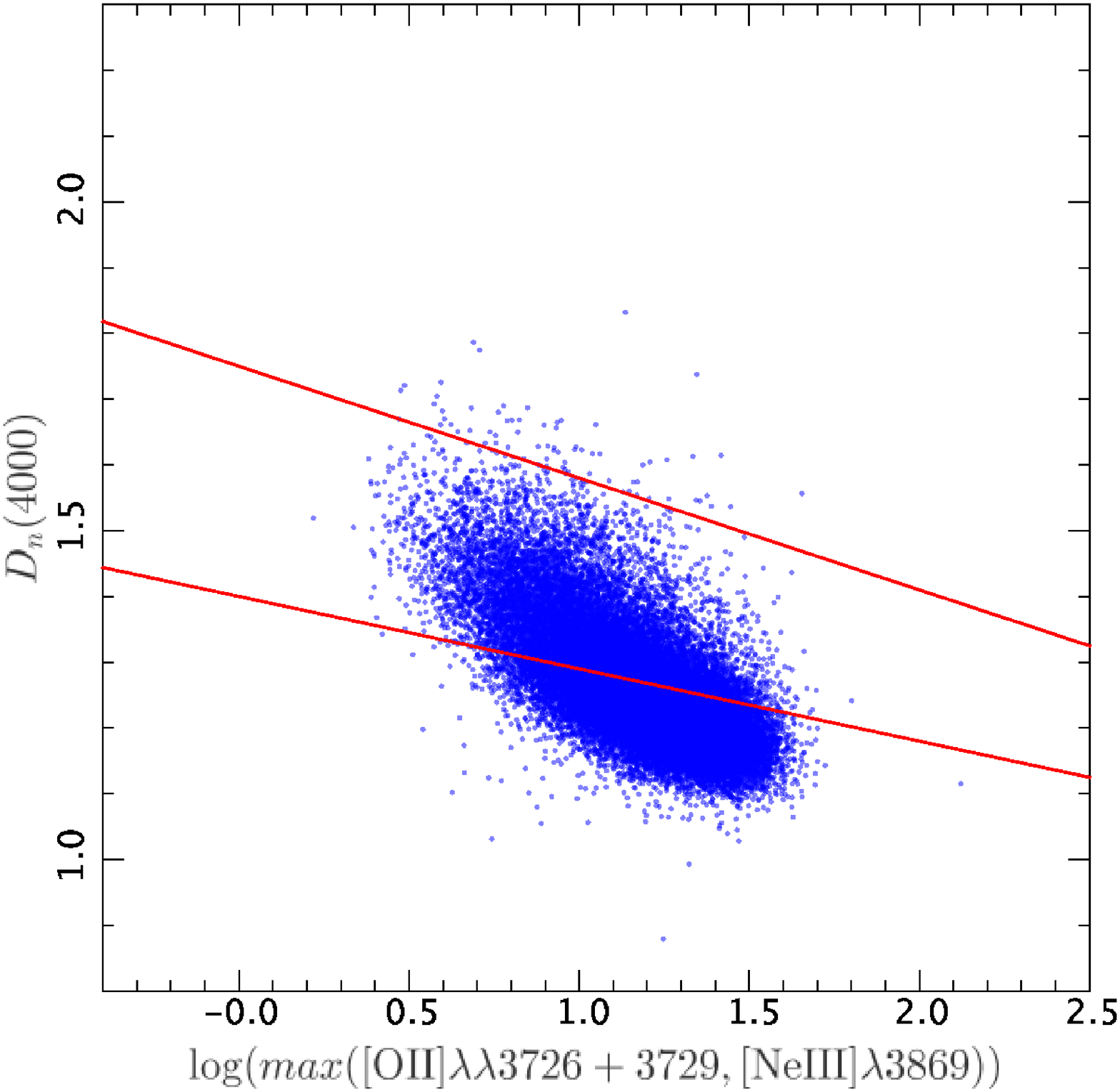}\includegraphics[width=0.29\paperwidth]{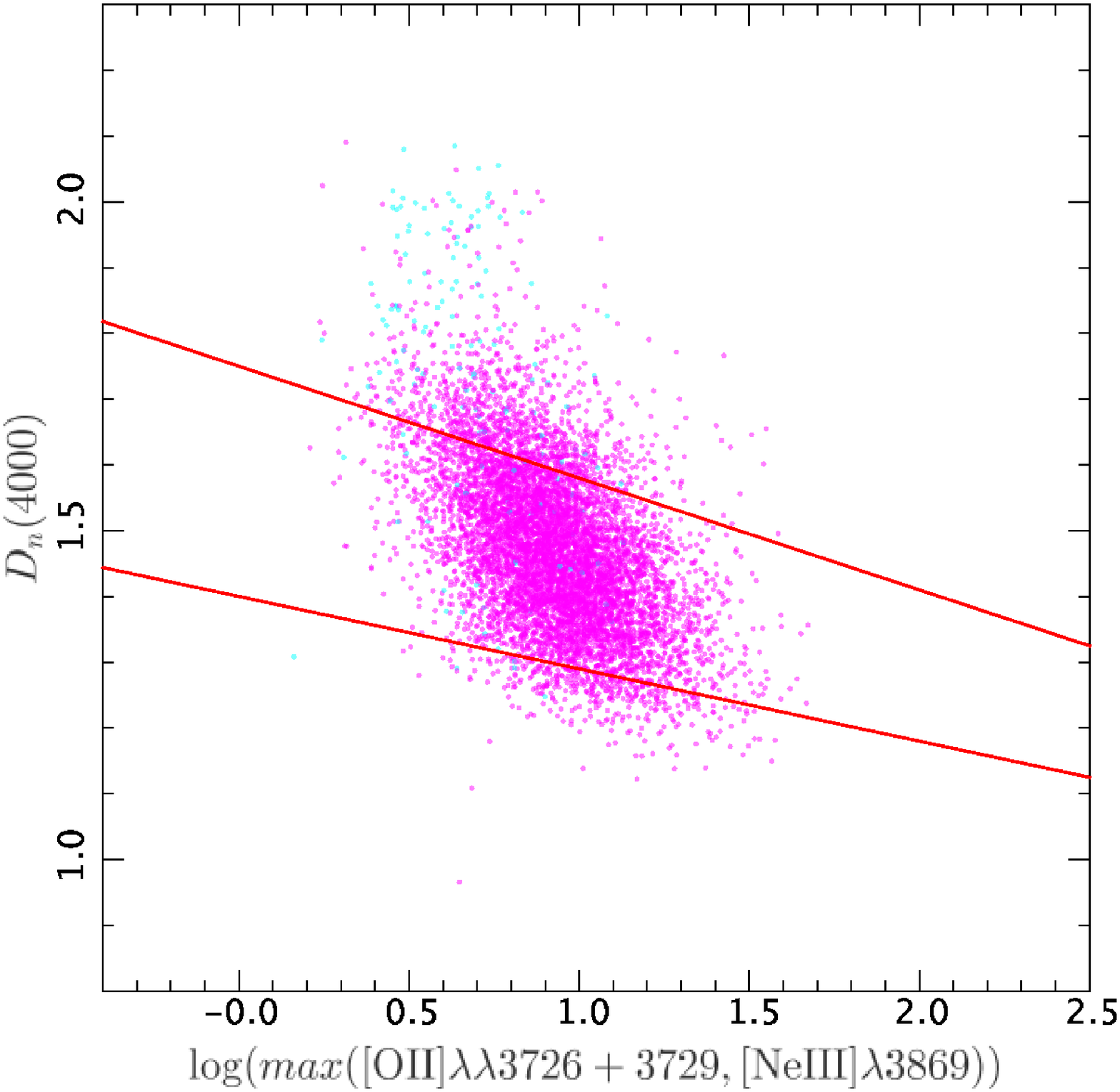}
\par\end{centering}

\begin{centering}
\includegraphics[width=0.29\paperwidth]{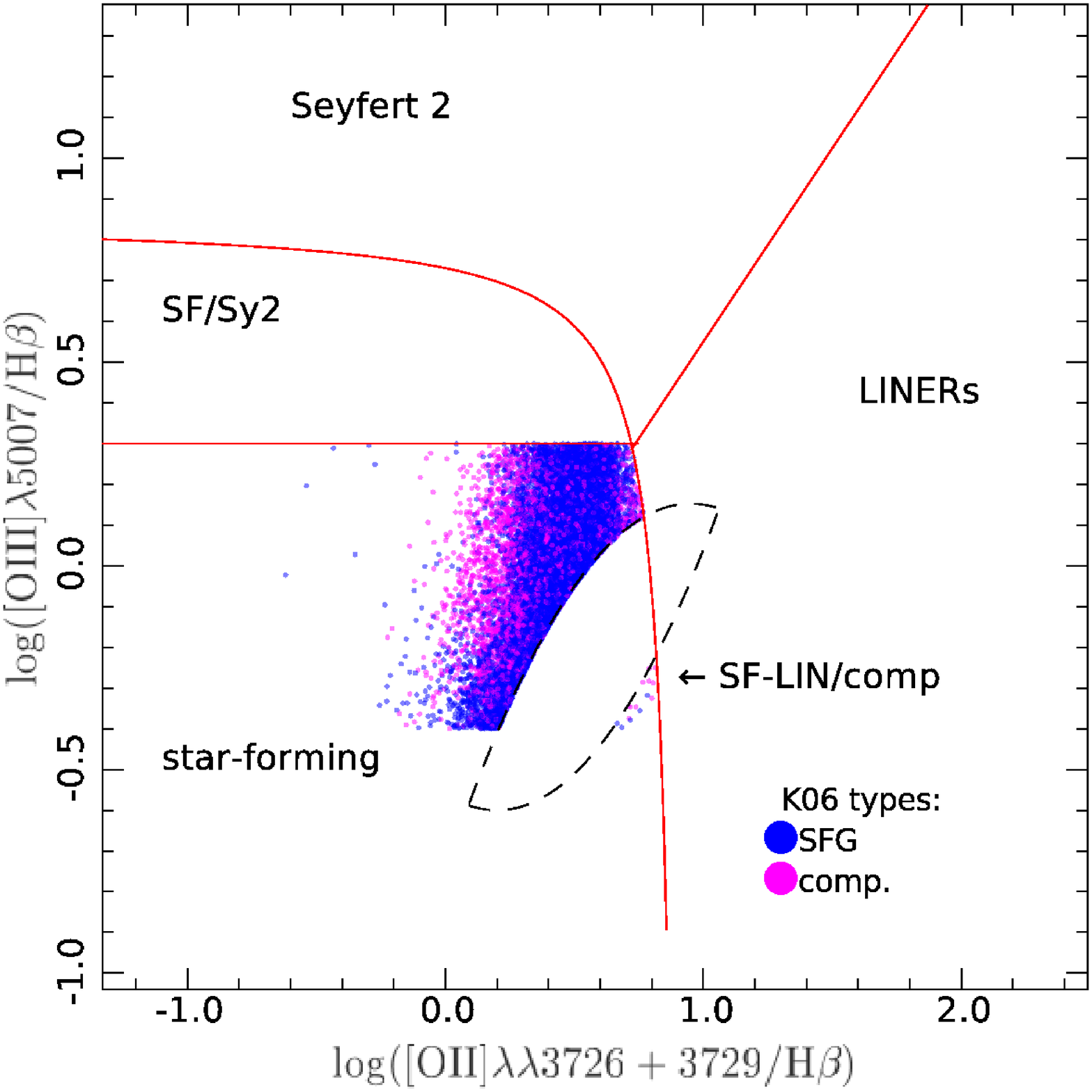}\includegraphics[width=0.29\paperwidth]{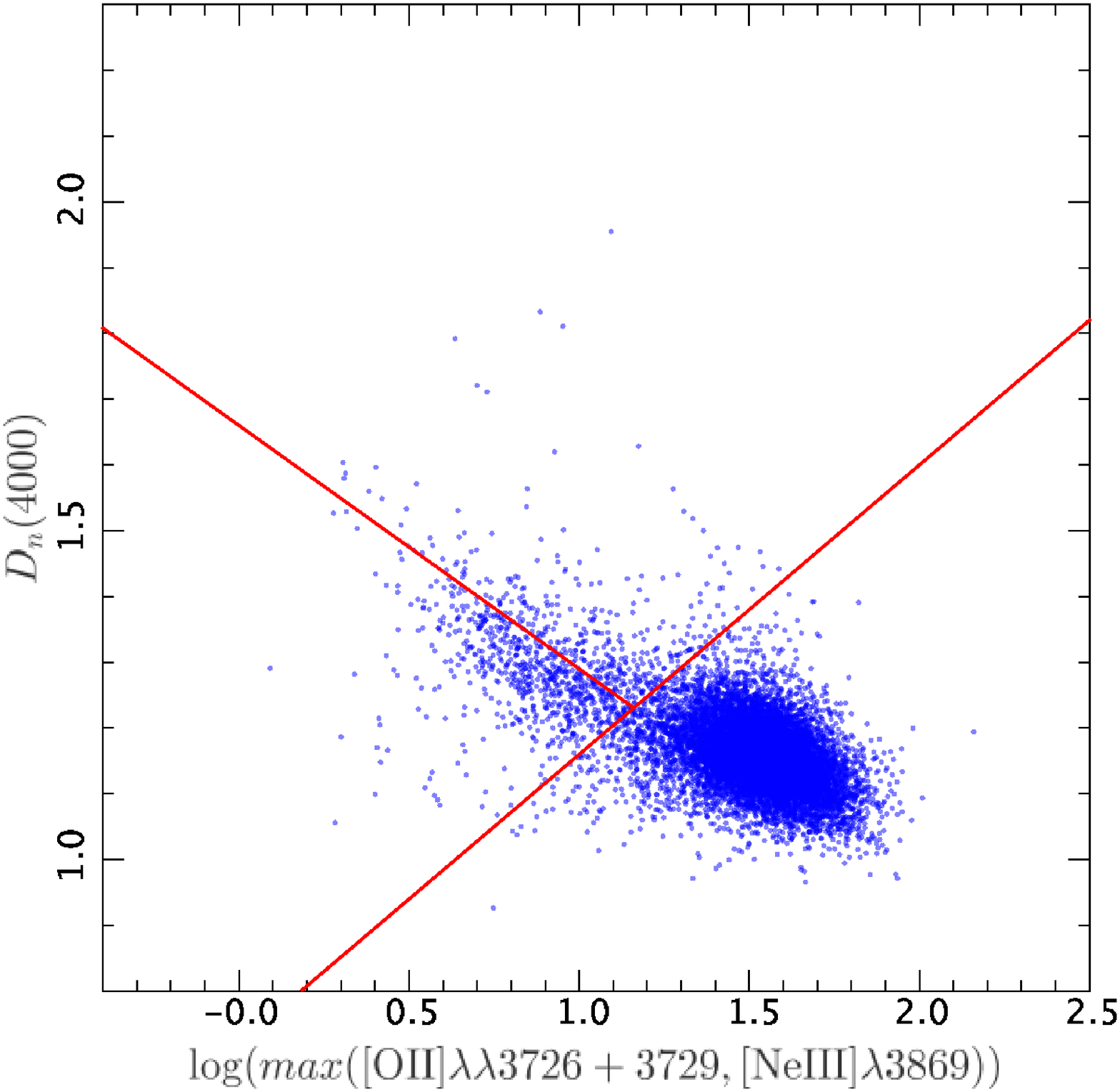}\includegraphics[width=0.29\paperwidth]{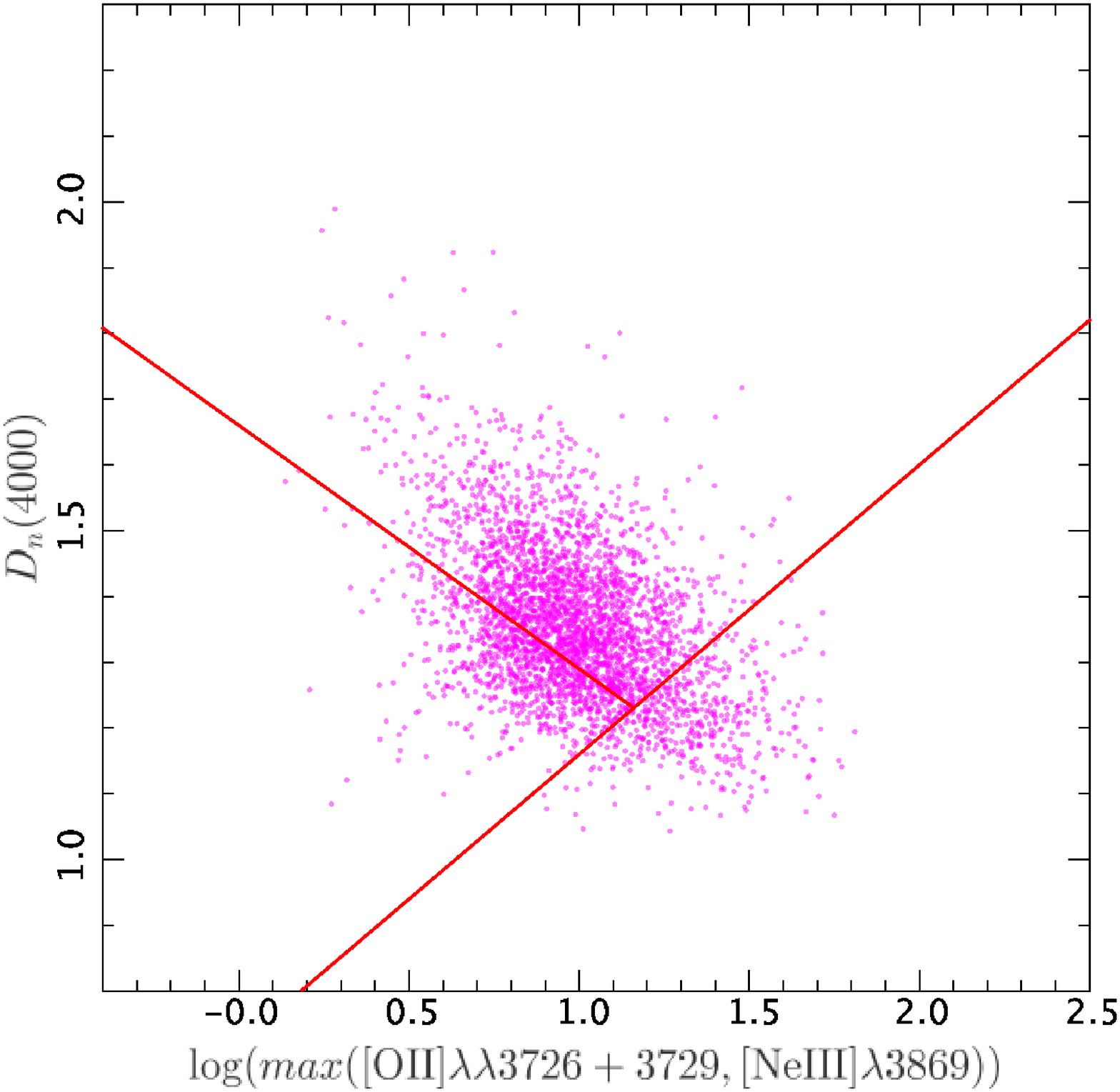}
\par\end{centering}

\begin{centering}
\includegraphics[width=0.29\paperwidth]{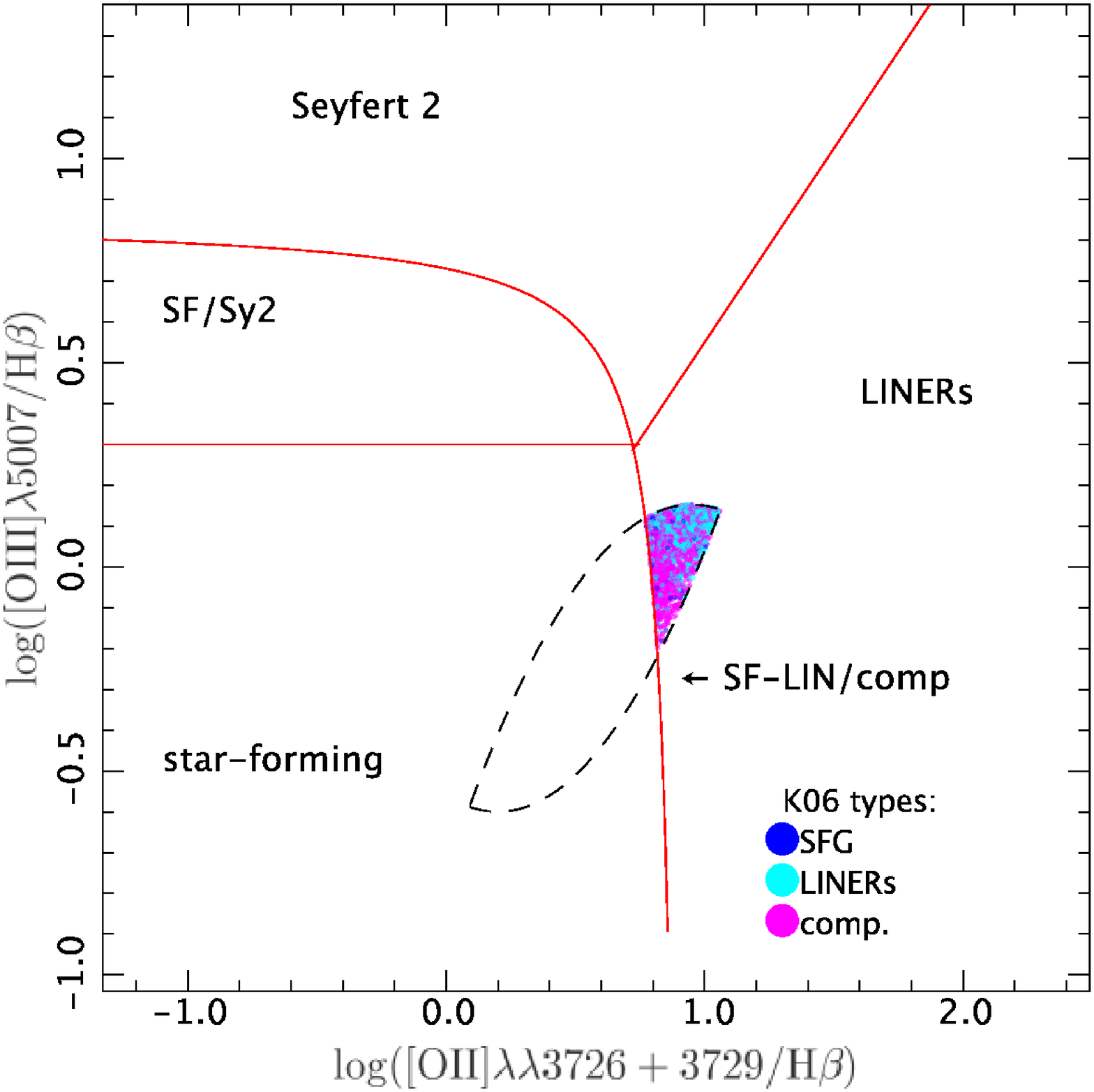}\includegraphics[width=0.29\paperwidth]{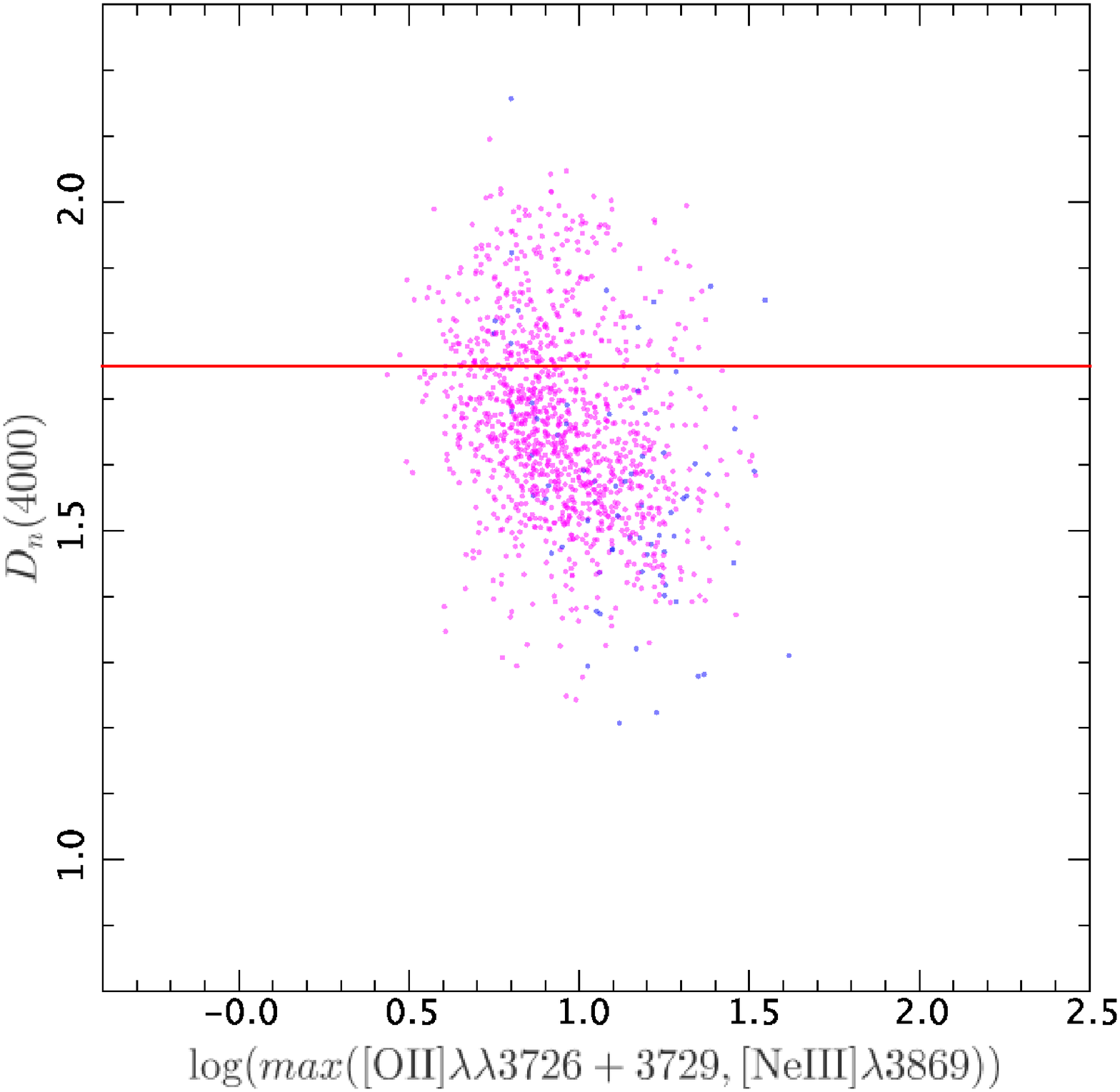}\includegraphics[width=0.29\paperwidth]{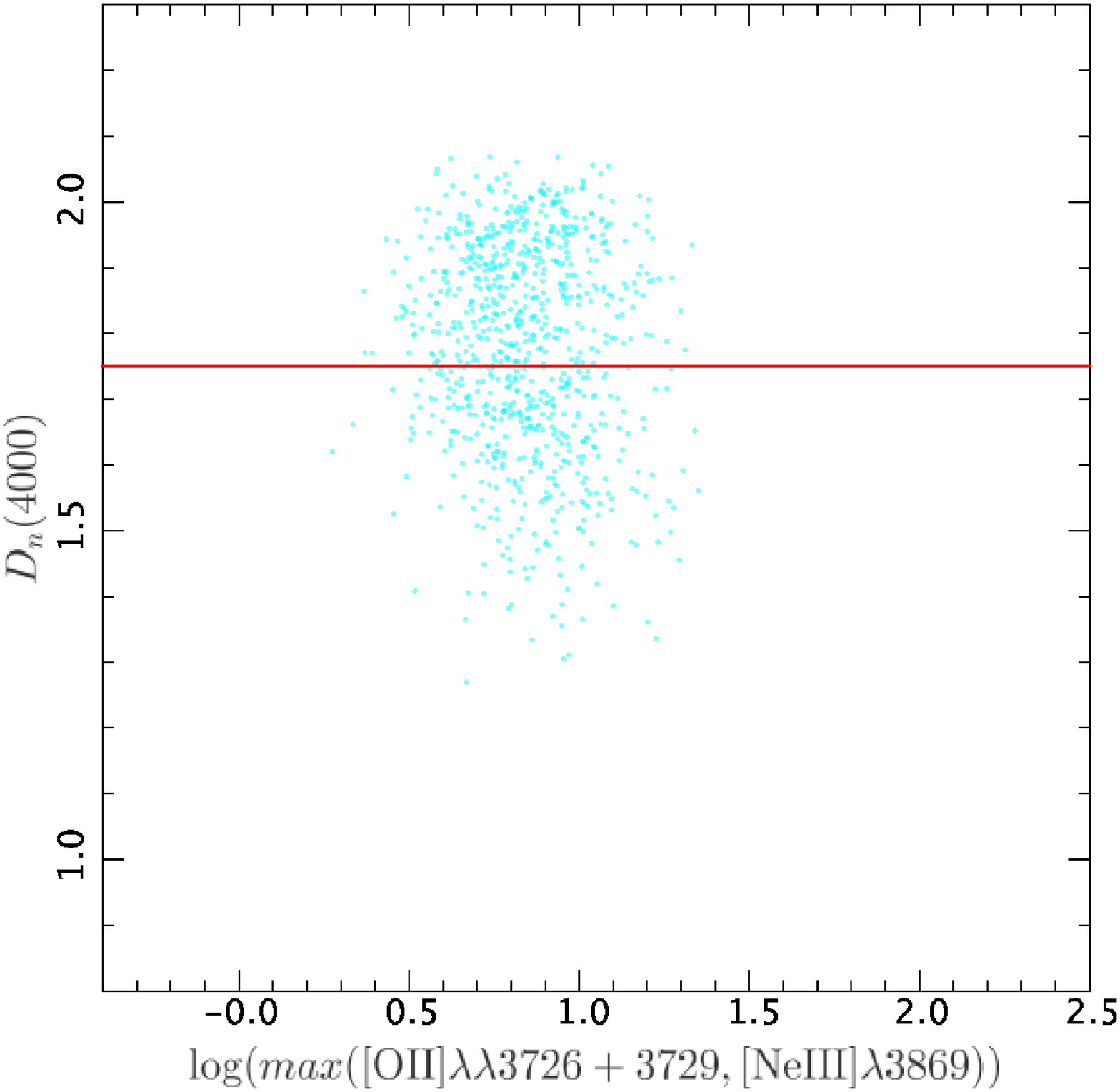}
\par\end{centering}

\caption{\emph{Left}: In the $\log\left([\mathrm{O}\textsc{iii}]\lambda5007/\mathrm{H}\beta\right)$
vs. $\log\left([\mathrm{O}\textrm{\textsc{ii}}]\lambda\lambda3726+3729/\mathrm{H}\beta\right)$
diagram, the different regions where galaxies of different types are
mixed (color-code according to the reference K06 diagnostic). The
curves are the ones from the \citet{2009lama} diagnostic. \emph{Center
and right}: The same points as in the associated left panels, but
now in the $D_{n}(4000)$ vs. $\mathrm{max\left(EW[O\textsc{ii}],EW[Ne\textsc{iii}]\right)}$
diagram. For clarity, we plot the different types in two panels.The
solid lines show the adopted demarcation lines (see text for details).
The dashed line in top-right panels is the separation adopted by \citet{2006MNRAS.371..972S}.}

\label{Flo: Mixed areas DEW & New blue}
\end{figure*}

Figure~\ref{Flo: Mixed areas DEW & New blue} shows four regions
where galaxies of different types (according to the reference K06diagnostic)
are confused in the $\log\left([\mathrm{O}\textsc{iii}]\lambda5007/\mathrm{H}\beta\right)$
vs. $\log\left([\mathrm{O}\textrm{\textsc{ii}}]\lambda\lambda3726+3729/\mathrm{H}\beta\right)$
diagram. It shows also in its center and right panels how these galaxies
behave in the $D_{n}(4000)$ vs. $\mathrm{max\left(EW[O\textsc{ii}],EW[Ne\textsc{iii}]\right)}$
diagram. From top to bottom, the four studied regions are SFG/Sy2,
SFG/comp, another SFG/comp region not defined in the L10 diagnostic
but where a non negligible number of the composites (25\%) are still
mixed with SFGs, and LIN/comp.

\subsection{The SFG/Sy2 region}

In the L10 SFG/Sy2 region (see Fig.~\ref{Flo: Mixed areas DEW & New blue}
top), K06 Seyfert 2 and SFGs are confused. It is unfortunately obvious
in the bottom-left panel that the DEW diagnostic\emph{ does not} separate
the two classes of objects correctly in the L10 SFG/Sy2 region. We
thus propose a new demarcation line to separate K06 Seyfert 2 from
SFGs, \emph{valid only in the L10 SFG/Sy2 region}, with the equation
\begin{equation}
D_{n}(4000)=0.22\times\mathrm{\log\left(max\left(EW[O\textsc{ii}],EW[Ne\textsc{iii}]\right)\right)}+0.97.\label{eq: Red dashed DEW}
\end{equation}
Seyfert 2 would fall above this line, SFG below.The slope and zero
point of this line have been optimized by minimizing on a grid the
following function:
\begin{equation}
\chi^{2}=(1-S^{A})^{2}+(1-S_{B})^{2}+(C^{A})^{2}+(C_{B})^{2},\label{eq:minimize}
\end{equation}
where $S^{A}$, $S_{B}$, $C^{A}$, $C_{B}$ are the success rate
for AGN above the defined line, the success rate for SFG below the
line, the contamination by SFG above the line, and the contamination
by AGN below the line (all values between 0 and 1), respectively.
Indeed, we want to maximize the success rates and minimize the contamination
at the same time above and below the defined line. The grid is done
in $0.01$dex steps in both slope and zero point. To minimize computer
time, limits on this grid are defined by eye.

\begin{table}
\caption{Success chart for the supplementary M11 diagnostic in the L10 SFG/Sy2
region.}

\begin{tabular}{ccc}
\hline 
\hline
 & \multicolumn{2}{c}{reference K06}\tabularnewline
\hline 
M11 & SFG & Seyfert 2\tabularnewline
\hline 
\emph{total} & \emph{100} & \emph{100}\tabularnewline
SFG & 99.10 & 3.10\tabularnewline
Seyfert 2 & 0.90 & 96.90\tabularnewline
\hline 
\end{tabular}

\label{Flo: succes SFG-Sey2}
\end{table}

We have established the success of our new diagnostic in the L10 SFG/Sy2
region (see Table~\ref{Flo: succes SFG-Sey2}). This chart shows
that our demarcation line works almost perfectly and can be used in
that area. We have now correctly classified almost all actual Seyfert
2 in our sample: 97\% of those in the L10 SFG/Sy2 region are classified
as Seyfert 2. Given that 59\% of the K06 Seyfert 2 in the whole sample
were already correctly classified, and 26\% of them classified as
L10 SFG/Sy2, this increases the global success rate to 85\%. This
is the best success rate one can obtain by combining L10 and DEW diagrams.
The contamination in the SFG/Sy2 region is made of 3.1\% SFGs above
the line defined by Eq.~\ref{eq: Red dashed DEW} and 0.9\% Seyfert
2 below it.

\subsection{The SFG/comp region}

The L10 SF/comp (see Fig.~\ref{Flo: Mixed areas DEW & New blue}
second line) contain composites, SFGs, and very few LINERs. Following
the optimization procedure explained above, we find the following
equation which separate as many SFGs as possible from composites:

\begin{equation}
D_{n}(4000)=-0.11\times\mathrm{\log\left(max\left(EW[O\textsc{ii}],EW[Ne\textsc{iii}]\right)\right)}+1.4,\label{eq: newmix2}
\end{equation}
where SFGs are below this line. Since SFGs dominate the sample, the
region below this line is composed of 98\% SFGs. Conversely, the region
above this line is still a mix between SFGs (59\%), composites (36\%),
and a few LINERs (4\%). We again applied the optimization procedure
but now only consider the latest region, in order to isolate pure
composites as much as possible. We obtain:
\begin{equation}
D_{n}(4000)=-0.17\times\mathrm{\log\left(max\left(EW[O\textsc{ii}],EW[Ne\textsc{iii}]\right)\right)}+1.75,\label{eq: newmix2b}
\end{equation}
where composites are above the line, SFG/comp below it (i.e. between
the two lines).

\begin{table}
\caption{Success chart for the supplementary M11 diagnostic in the L10 SFG/comp
region.}

\begin{tabular}{cccc}
\hline 
\hline
 & \multicolumn{3}{c}{reference K06}\tabularnewline
\hline 
M11 & SFG & LINERs & composites\tabularnewline
\hline 
\emph{total} & \emph{100} & \emph{100} & \emph{100}\tabularnewline
SFG & 63.69 & 3.47 & 5.31\tabularnewline
composites & 0.19 & 70.14 & 13.95\tabularnewline
SFG/comp & 36.12 & 26.39 & 80.74\tabularnewline
\hline 
\end{tabular}

\label{Flo: succes SFGLINcomp}
\end{table}
\begin{table}
\caption{Contamination chart for the supplementary M11 diagnostic in the L10
SFG/comp region.}

\begin{tabular}{ccccc}
\hline 
\hline
 & \multicolumn{4}{c}{reference K06}\tabularnewline
\hline 
M11 & \emph{total} & SFG & LINERs & composites\tabularnewline
\hline 
SFG & \emph{100} & 98.30 & 0.02 & 1.67\tabularnewline
composites & \emph{100} & 5.85 & 8.21 & 85.93\tabularnewline
SFG/comp & \emph{100} & 68.81 & 0.19 & 31.00\tabularnewline
\hline 
\end{tabular}

\label{Flo: conta SFGLINcomp}
\end{table}

Tables~\ref{Flo: succes SFGLINcomp} and~\ref{Flo: conta SFGLINcomp}show
the success chart and the contamination chart of the supplementary
M11 diagnostic, in the L10 SFG/comp region. From all the K06 SFGs
present in this region, 64\% are now correctly classified as SFGs,
another 36\% being still ambiguously classified as SFG/comp.\emph{It
is unfortunately impossible }to increase this success rate without
misclassifying too many K06 composites as SFGs. Only 14\% of K06 composites
could be isolated. Newly isolated SFGs are not significantly contaminated
by K06 composites (2\%), and neither are the newly isolated composites
by K06 SFGs (6\%). The majority of the K06 of composites (81\%) are
still ambiguously classified as SFG/comp. The best reason to use this
diagram is clearly for isolating SFGs. The SFG/comp region is now
made of 69\% K06 SFGs and 31\% K06 composites, which is more balanced
than in Paper~I (respectively 83\% and 17\%).

\subsection{Additional region of composites mixed with SFGs}

We define a last region where K06 composites are mixed with K06 SFGs
in the L10 diagnostic. It is located below the SFG/Sy2 region, and
above $\log\left([\mathrm{O}\textsc{iii}]\lambda5007/\mathrm{H}\beta\right)>-0.4$,
excluding the SFG/comp region (see Fig.~\ref{Flo: Mixed areas DEW & New blue}
third line). Even though we can see that these K06 SFGs are mostly
spread over the bottom right of the DEW diagnostic diagram and K06
composites are slightly above, there is still a small area where they
get together.

Following the same optimization procedure as above, we first fit the
optimized demarcation line between pure SFGs and SFGs mixed with composites:
\begin{equation}
D_{n}(4000)=0.44\times\mathrm{\log\left(max\left(EW[O\textsc{ii}],EW[Ne\textsc{iii}]\right)\right)}+0.72,\label{eq: newmix3below}
\end{equation}
where SFGs are below this line. As in previous section, the region
above this line contains a mix of SFGs (25\%) and composites (75\%).
In this region, we fit another optimized demarcation line between
pure composites and composites mixed with SFGs:
\begin{equation}
D_{n}(4000)=-0.37\times\mathrm{\log\left(max\left(EW[O\textsc{ii}],EW[Ne\textsc{iii}]\right)\right)}+1.66,\label{eq: newmix3above}
\end{equation}
where composites are above this line, and SFG/comp below it (i.e.
between the two lines). We finally add the remaining mixed galaxies
to the SFG/comp type.

\begin{table}
\caption{Success chart for the supplementary M11 diagnostic in the additional
region of the L10 diagnostic where SFG are mixed with composites.}

\begin{tabular}{ccc}
\hline 
\hline
 & \multicolumn{2}{c}{reference K06}\tabularnewline
\hline 
M11 & SFG & composites\tabularnewline
\hline 
\emph{total} & \emph{100} & \emph{100}\tabularnewline
SFG & 92.10 & 12.69\tabularnewline
composites & 2.90 & 61.97\tabularnewline
SFG/comp & 5.00 & 25.34\tabularnewline
\hline 
\end{tabular}

\label{Flo: successSFGcomp}
\end{table}

\begin{table}
\caption{Contamination chart for the supplementary M11 diagnostic in the additional
region of the L10 diagnostic where SFG are mixed with composites.}

\begin{tabular}{cccc}
\hline 
\hline
 & \multicolumn{3}{c}{reference K06}\tabularnewline
\hline 
M11 & \emph{total} & SFG & composites\tabularnewline
\hline 
SFG & \emph{100} & 96.39 & 3.61\tabularnewline
composites & \emph{100} & 14.50 & 85.30\tabularnewline
SFG/comp & \emph{100} & 42.08 & 57.92\tabularnewline
\hline 
\end{tabular}

\label{Flo: contSFGcomp}
\end{table}

Tables~\ref{Flo: succes SFGLINcomp} and~\ref{Flo: contSFGcomp}
show the success and contamination charts of the supplementary M11
diagnostic in this last region. Of all K06 SFGs present in this region,
92\% are now correctly classified, and another 5\% are ambiguously
classified as SFG/comp. We also managed to isolate 62\% of the K06
composites falling in this region, which is an improvement compared
to the L10 diagnostic. Conversely, 12\% of the K06 composites are
now unfortunately misclassified as SFGs. This is still an improvement
: we recall that 100\% of the K06 composites in this region used to
be classified as SFGs with the L10 diagnostic. Most of them (62\%)
are now classified as composite galaxies, and another 25\% are still
ambiguously classified as SFG/comp. The composites defined in this
region are contaminated by 15\% K06 SFGs, which may not be neglected.
The SFG/comp galaxies defined in this region are made of approximately
half K06 SFGs (42\%) and half K06 composites (58\%).

\subsection{The LIN/comp region}

Figure~\ref{Flo: Mixed areas DEW & New blue} (bottom) shows the
LIN/comp region of the L10 diagnostic in the DEW diagnostic diagram.
It is clear that this diagram cannot be used to isolate LINERs cleanly
from composites. Indeed, one may argue that LINERs are more concentrated
on the top and composites on the bottom of the diagram, so we propose
a straight line at $D_{n}(4000)\approx1.75$ as a separation. Using
this line we find, however, 61\% LINERs and 39\% composites above,
28\% and 72\% below respectively, which is unsatisfactory. Thus, we
do not update the LIN/comp region as in Paper~I.

\subsection{Discussion}

\begin{figure*}
\begin{centering}
\includegraphics[width=0.3\paperwidth]{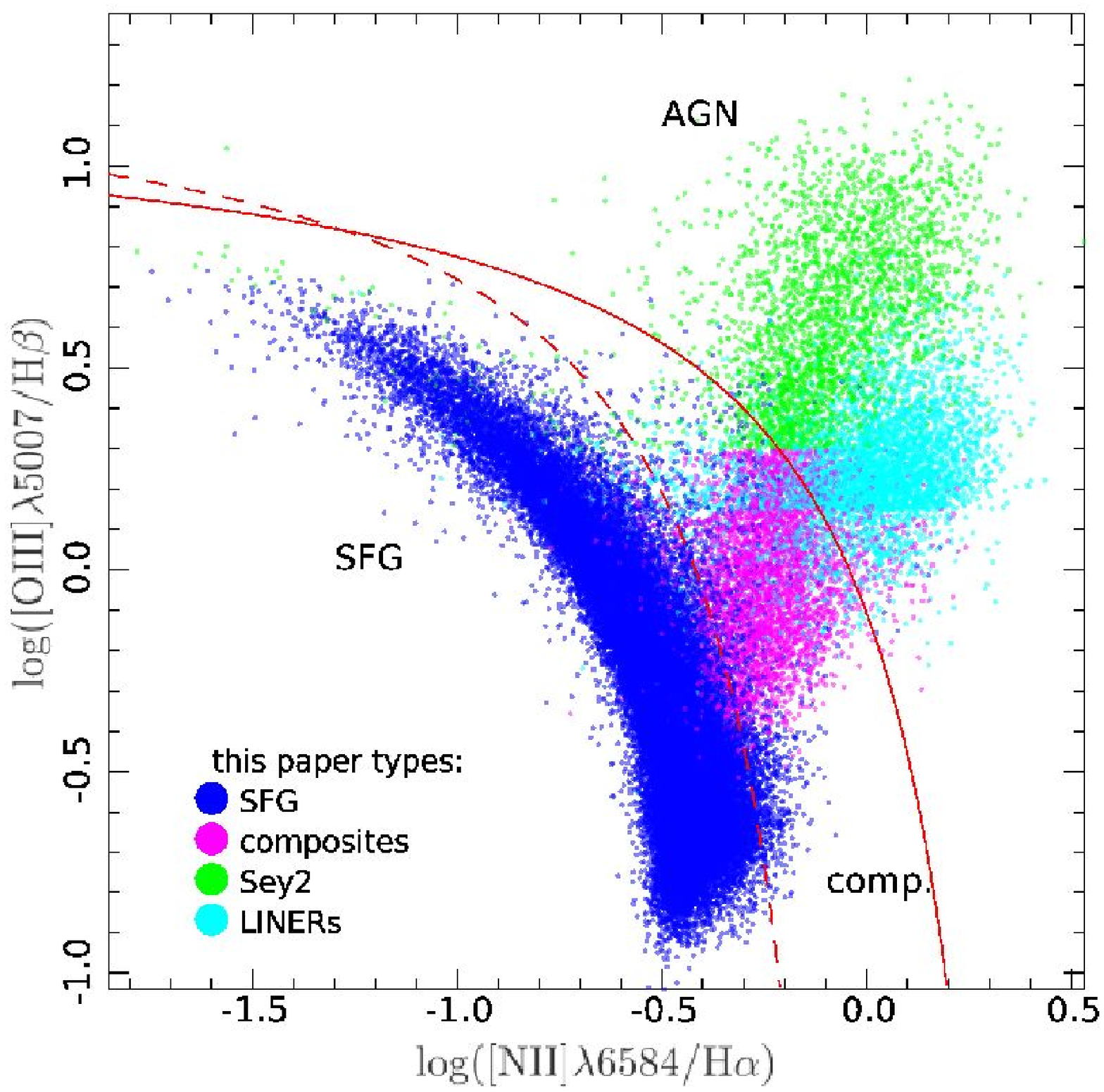} \includegraphics[width=0.3\paperwidth]{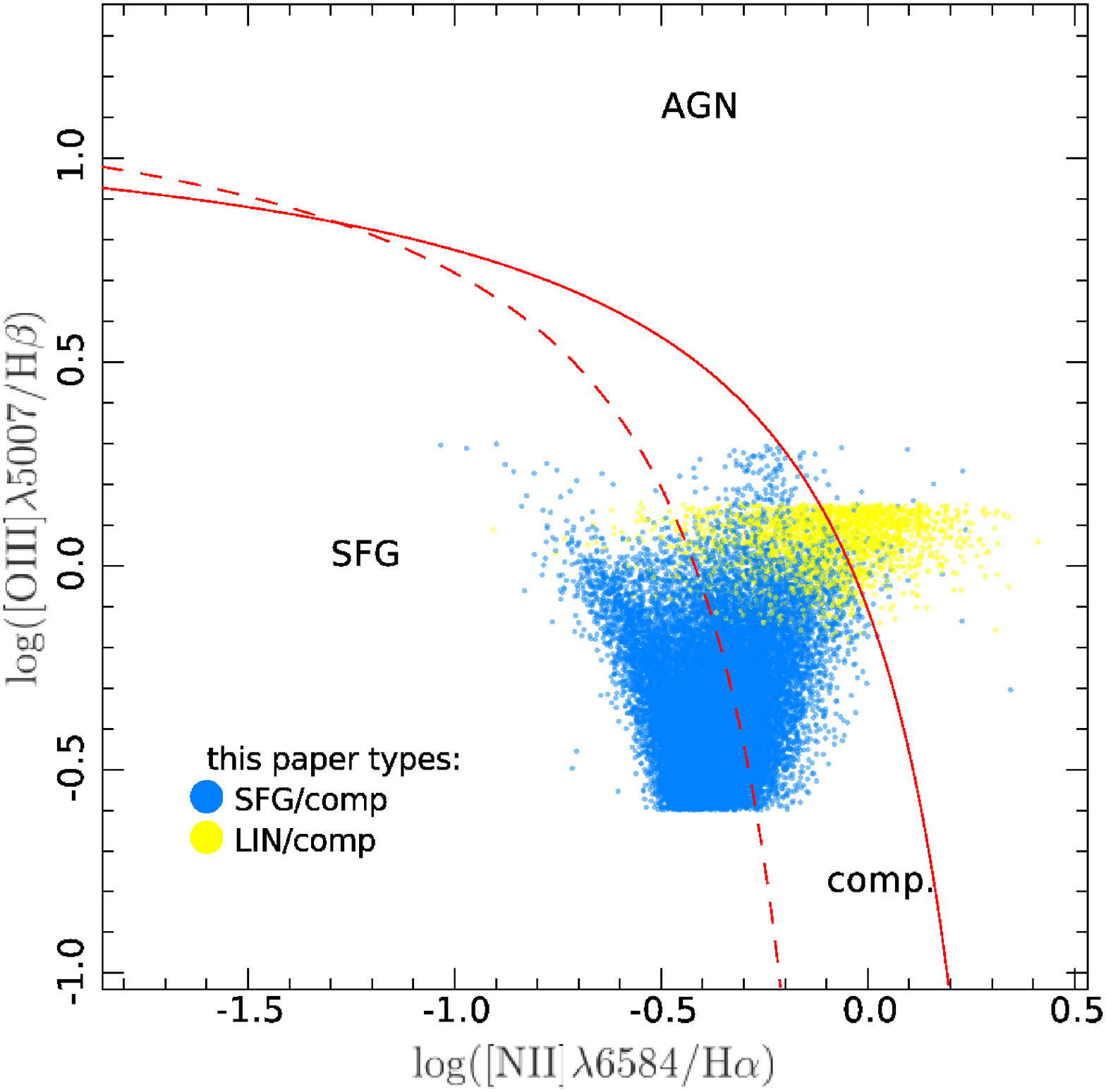}
\par\end{centering}

\caption{Results of the new diagnostic derived in the present paper in the
standard log$\left([\mathrm{O\textrm{\textsc{iii}}}]\lambda5007/\mathrm{H}\beta\right)$
vs. log$\left([\mathrm{N}\textsc{ii}]\lambda6583/\mathrm{H}\alpha\right)$
diagram. For clarity, only SFGs, AGNs and composites are shown in
the left panel, while SFG/comp and LIN/comp are shown in the right
panel. Same color code as in Fig.~\ref{Flo:summary}.}

\label{Flo:summary-1}
\end{figure*}

Figure~\ref{Flo:summary-1} shows our supplementary M11 diagnostic
combined with L10 diagnostic, in one of the standard K06 diagnostic
diagrams. The left panel looks pretty good: SFGs Seyfert 2, LINERs,
and some composites lie almost perfectly in the correct corresponding
regions of this diagram. In contrast, the right hand panel shows the
limitations: objects that are still ambiguously classified or as SFG/comp
or as LIN/comp. Anyway, comparing Fig.~\ref{Flo:summary-1} to the
bottom left hand panel of Fig.~\ref{Flo:summary}, we see a clear
improvement.

\begin{table}
\caption{Overall success chart for the L10 and M11 diagnostics.}

\begin{tabular}{ccccc}
\hline 
\hline
 & \multicolumn{4}{c}{reference K06}\tabularnewline
\hline 
L10/M11 & SFG & Composites & Seyfert 2 & LINERs\tabularnewline
\hline 
\emph{total} & \emph{100} & \emph{100} & \emph{100} & \emph{100}\tabularnewline
SFG & 77.97 & 8.92 & 0.92 & 0.10\tabularnewline
SFG/comp & 20.99 & 50.97 & 0.47 & 0.98\tabularnewline
composites & 0.66 & 23.44 & 5.46 & 4.30\tabularnewline
Seyfert 2 & 0.12 & 0.93 & 84.71 & 3.93\tabularnewline
LINERs & 0.17 & 6.94 & 8.00 & 73.51\tabularnewline
LIN/comp & 0.11 & 8.81 & 0.44 & 17.18\tabularnewline
\hline 
\emph{total SFG}{\small $^{1}$} & \emph{98.96} & \emph{59.88} & \emph{1.39} & \emph{1.08}\tabularnewline
\emph{total LINERs}{\small $^{2}$} & \emph{0.27} & \emph{15.74} & \emph{8.44} & \emph{90.70}\tabularnewline
\emph{total comp.}{\small $^{3}$} & \emph{21.75} & \emph{83.22} & \emph{6.38} & \emph{22.46}\tabularnewline
\hline 
\end{tabular}

{\small $^{1}$ SFG+SFG/com}{\small \par}

{\small $^{2}$ LINERs+LIN/comp}{\small \par}

{\small $^{3}$ composites+SFG/comp+LIN/comp.}{\small \par}

\label{Flo: Success chart Hybrid}
\end{table}

\begin{table}
\caption{Overall contamination chart for the L10 and M11 diagnostics.}

\begin{tabular}{cccccc}
\hline 
\hline
 & \multicolumn{5}{c}{reference K06}\tabularnewline
\hline 
L10/M11 & \emph{total} & SFG & Composites & Seyfert 2 & LINERs\tabularnewline
\hline 
SFG & \emph{100} & 97.68 & 2.26 & 0.05 & 0.01\tabularnewline
SFG/comp & \emph{100} & 66.81 & 32.89 & 0.07 & 0.23\tabularnewline
composites & \emph{100} & 11.02 & 79.77 & 3.99 & 5.23\tabularnewline
Seyfert 2 & \emph{100} & 2.73 & 4.42 & 86.20 & 6.66\tabularnewline
LINERs & \emph{100} & 2.28 & 19.40 & 4.80 & 73.51\tabularnewline
LIN/comp & \emph{100} & 3.37 & 56.57 & 0.61 & 39.46\tabularnewline
\hline 
\emph{total SFG}{\small $^{1}$} & \emph{100} & \emph{75.83} & \emph{15.37} & \emph{3.30} & \emph{5.50}\tabularnewline
\emph{total LINERs}{\small $^{2}$} & \emph{100} & \emph{2.61} & \emph{30.68} & \emph{3.53} & \emph{63.18}\tabularnewline
\emph{total comp.}{\small $^{3}$} & \emph{100} & \emph{53.67} & \emph{41.63} & \emph{0.68} & \emph{4.02}\tabularnewline
\hline 
\end{tabular}

{\small $^{1}$ SFG+SFG/com}{\small \par}

{\small $^{2}$ LINERs+LIN/comp}{\small \par}

{\small $^{3}$ composites+SFG/comp+LIN/comp.}{\small \par}

\label{Flo: Contamination chart Hybrid}
\end{table}

Tables~\ref{Flo: Success chart Hybrid} and~\ref{Flo: Contamination chart Hybrid}
establish new contamination and success charts in order to get a more
precise measurement of this improvement. Success chart shows that
1\% of K06 SFG are classified in a region where they are not supposed
to be. K06 composites are predominantly found in the SFG/comp region
(51\%) than in the composite region (23\%). This is an improvement
over Paper~I, but it shows that neither the L10 nor the DEW diagrams
are really good at identifying composites at high redshift. K06 Seyfert
2 galaxies and LINERs have a high success rate (respectively 85\%
and 74\%), which is for Seyfert 2 galaxies a really good improvement
compared to Paper~I. For K06 LINERs, the success rate increases to
91\% including the LIN/comp, but one has to be aware that this region
is actually made of only 39\% K06 LINERs and is dominated by 57\%
K06 composites. 

Nevertheless, one can conclude from the contamination chart that SFGs,
Seyfert 2, and LINERs are not significantly contaminating each other.
The main contamination comes in all cases from composites: 33\% in
the SFG/comp region, 19\% in the LINERs region, and 57\% in the LIN/comp
region. It is conversely very low in the SFG and Seyfert 2 regions.

One could worry about aperture effects. Indeed, SDSS spectra are based
on 3'' fibers. This may end in overestimated $D_{n}(4000)$ values
for close objects where only the central bulge is covered by the fiber.
However, one can see in Fig.~\ref{Flo: Mixed areas DEW & New blue}
that only objects with $D_{n}(4000)<1.5$ could change their classification
with a significant aperture effect. Looking at Fig.~16 in \citet{2003MNRAS.341...33K},
we see that those objects do not actually suffer from a strong aperture
effect. We conclude that our diagnostic is not biased by this effect.

Finally, we invite the reader to take a look at the right hand panels
of Fig.~\ref{Flo: AGN count newblue}. It shows the AGN counts obtained
with our new supplementary M11 diagnostic combined with L10 diagnostic.
We clearly see that the diagnostic derived in the present paper is
the one that follows the reference K06 curve more accurately. There
are still problems for $\log\left([\mathrm{O}\textsc{iii}]/\mathrm{H}\beta\right)\lesssim0.25$,
which is normal since we did not manage to change the classification
of LINERs by adding the DEW diagnostic diagram.

\section{Conclusion \label{sec:Conclusion}}

By adding the M11 diagnostic to the L10 diagnostic derived in Paper~I,
we now have a very good classification of emission-line galaxies that
can be used on high-redshifts samples. The main improvements compared
to Paper~I are
\begin{itemize}
\item The unambiguous classification of objects in the former SFG/Sy2 region
as SFGs or Seyfert 2.
\item The unambiguous classification of some of the objects in the SFG/comp
region as SFGs or composites (where no composites at all were found
in Paper~I).
\item A better definition of the SFG/comp region, which leaves fewer possible
composites not flagged as such. We emphasize again that this region
is in any case dominated by SFGs.
\end{itemize}
No improvements could have been done in the LIN/comp region, which
is left unchanged compared to Paper~I.

In order to use the diagnostic derived in this paper, one should follow
these steps.
\begin{enumerate}
\item Classify objects in the $\log\left([\mathrm{O}\textsc{iii}]\lambda5007/\mathrm{H}\beta\right)$
vs. $\log\left([\mathrm{O}\textrm{\textsc{ii}}]\lambda\lambda3726+3729/\mathrm{H}\beta\right)$
with the L10 diagnostic derived in Paper~I (see also equations in
Sect.~\ref{sub:L10-classification}).
\item Classify objects falling in the SFG/Sy2 region as SFGs or Seyfert
2 using Eq.~\ref{eq: Red dashed DEW}.
\item Isolate objects falling the the SFG/comp region as

\begin{enumerate}
\item SFGs using Eq.~\ref{eq: newmix2}, and
\item composites using Eq.~\ref{eq: newmix2b}.
\end{enumerate}
\item Define a new SFG/comp region inside the SFG region using $\log\left([\mathrm{O}\textsc{iii}]\lambda5007/\mathrm{H}\beta\right)>-0.4$,
so inside this new region, isolate

\begin{enumerate}
\item SFGs using Eq.~\ref{eq: newmix3below}, and
\item composites using Eq.~\ref{eq: newmix3above}.
\end{enumerate}
\end{enumerate}
In both points (3) and (4) above, objects not classified as SFGs or
composites remain of the ambiguous SFG/comp type. We invite the reader
to look at the {}``JClassif'' software (available at: \url{http://www.ast.obs-mip.fr/galaxie/}),
which performs these steps automatically on any sample, as well as
other classification schemes.

Table~\ref{Flo: Contamination chart Hybrid} may be used as a probability
chart showing whether each type in our diagnostic is one of the K06
reference types. However, we warn the reader that the relative proportions
of SFGs, composites and AGNs in each regions of our diagnostic diagrams
may evolve with redshift compared to the SDSS sample. Finally, we
note that it is possible to upgrade this classification to higher
redshifts where $[\mathrm{O}\textsc{iii}]\lambda5007$ and $\mathrm{H}\beta$
emission lines get red-shifted out of the spectra ($1.0\apprle z\apprle1.5$
on optical spectra). To that goal one may use the equations provided
by \citet{2007MNRAS.381..125P}, which convert $[\mathrm{Ne}\textsc{iii}]\lambda3869$
and $\mathrm{H}\delta$ to $[\mathrm{O}\textsc{iii}]\lambda5007$
and $\mathrm{H}\beta$.
\begin{acknowledgements}
F.L. thanks {}``La cité de l'espace'' for financial support while
this paper was being written. The data used in this paper were produced
by a collaboration of researchers (currently or formerly) from the
MPA and the JHU. The team is made up of Stéphane Charlot (IAP), Guinevere
Kauffmann and Simon White (MPA), Tim Heckman (JHU), Christy Tremonti
(Max-Planck for Astronomy, Heidelberg - formerly JHU), and Jarle Brinchmann
(Sterrewach Leiden - formerly MPA). All data presented in this paper
were processed with the JClassif software, part of the Galaxie pipeline.
We thank the referee for useful corrections and suggestions for improving
the paper.

Funding for the SDSS and SDSS-II has been provided by the Alfred P.
Sloan Foundation, the Participating Institutions, the National Science
Foundation, the U.S. Department of Energy, the National Aeronautics
and Space Administration, the Japanese Monbukagakusho, the Max Planck
Society, and the Higher Education Funding Council for England. The
SDSS Web Site is http://www.sdss.org/.

The SDSS is managed by the Astrophysical Research Consortium for the
Participating Institutions. The Participating Institutions are the
American Museum of Natural History, Astrophysical Institute Potsdam,
University of Basel, University of Cambridge, Case Western Reserve
University, University of Chicago, Drexel University, Fermilab, the
Institute for Advanced Study, the Japan Participation Group, Johns
Hopkins University, the Joint Institute for Nuclear Astrophysics,
the Kavli Institute for Particle Astrophysics and Cosmology, the Korean
Scientist Group, the Chinese Academy of Sciences (LAMOST), Los Alamos
National Laboratory, the Max-Planck-Institute for Astronomy (MPIA),
the Max-Planck-Institute for Astrophysics (MPA), New Mexico State
University, Ohio State University, University of Pittsburgh, University
of Portsmouth, Princeton University, the United States Naval Observatory,
and the University of Washington. 
\end{acknowledgements}
\bibliographystyle{aa}
\addcontentsline{toc}{section}{\refname}\bibliography{bib}

\begin{thebibliography}{42}
\expandafter\ifx\csname natexlab\endcsname\relax\def\natexlab#1{#1}\fi

\bibitem[{{Abazajian} {et~al.}(2009){Abazajian}, {Adelman-McCarthy},
  {Ag{\"u}eros}, {Allam}, {Allende Prieto}, {An}, {Anderson}, {Anderson},
  {Annis}, {Bahcall}, \& et~al.}]{2009ApJS..182..543A}
{Abazajian}, K.~N., {Adelman-McCarthy}, J.~K., {Ag{\"u}eros}, M.~A., {et~al.}
  2009, \apjs, 182, 543

\bibitem[{{Abraham} {et~al.}(2004){Abraham}, {Glazebrook}, {McCarthy},
  {Crampton}, {Murowinski}, {J{\o}rgensen}, {Roth}, {Hook}, {Savaglio}, {Chen},
  {Marzke}, \& {Carlberg}}]{2004AJ....127.2455A}
{Abraham}, R.~G., {Glazebrook}, K., {McCarthy}, P.~J., {et~al.} 2004, \aj, 127,
  2455

\bibitem[{{Bacon} {et~al.}(2010){Bacon}, {Accardo}, {Adjali}, {Anwand},
  {Bauer}, {Biswas}, {Blaizot}, {Boudon}, {Brau-Nogue}, {Brinchmann},
  {Caillier}, {Capoani}, {Carollo}, {Contini}, {Couderc}, {Daguis{\'e}},
  {Deiries}, {Delabre}, {Dreizler}, {Dubois}, {Dupieux}, {Dupuy}, {Emsellem},
  {Fechner}, {Fleischmann}, {Fran{\c c}ois}, {Gallou}, {Gharsa}, {Glindemann},
  {Gojak}, {Guiderdoni}, {Hansali}, {Hahn}, {Jarno}, {Kelz}, {Koehler},
  {Kosmalski}, {Laurent}, {Le Floch}, {Lilly}, {Lizon}, {Loupias}, {Manescau},
  {Monstein}, {Nicklas}, {Olaya}, {Pares}, {Pasquini}, {P{\'e}contal-Rousset},
  {Pell{\'o}}, {Petit}, {Popow}, {Reiss}, {Remillieux}, {Renault}, {Roth},
  {Rupprecht}, {Serre}, {Schaye}, {Soucail}, {Steinmetz}, {Streicher}, {Stuik},
  {Valentin}, {Vernet}, {Weilbacher}, {Wisotzki}, \&
  {Yerle}}]{2010SPIE.7735E...7B}
{Bacon}, R., {Accardo}, M., {Adjali}, L., {et~al.} 2010, SPIE, 7735

\bibitem[{{Baldwin} {et~al.}(1981){Baldwin}, {Phillips}, \&
  {Terlevich}}]{1981PASP...93....5B}
{Baldwin}, J.~A., {Phillips}, M.~M., \& {Terlevich}, R. 1981, \pasp, 93, 5

\bibitem[{{Balestra} {et~al.}(2010){Balestra}, {Mainieri}, {Popesso},
  {Dickinson}, {Nonino}, {Rosati}, {Teimoorinia}, {Vanzella}, {Cristiani},
  {Cesarsky}, {Fosbury}, {Kuntschner}, \& {Rettura}}]{2010A&A...512A..12B}
{Balestra}, I., {Mainieri}, V., {Popesso}, P., {et~al.} 2010, \aap, 512, A12+

\bibitem[{{Balogh} {et~al.}(1999){Balogh}, {Morris}, {Yee}, {Carlberg}, \&
  {Ellingson}}]{1999ApJ...527...54B}
{Balogh}, M.~L., {Morris}, S.~L., {Yee}, H.~K.~C., {Carlberg}, R.~G., \&
  {Ellingson}, E. 1999, \apj, 527, 54

\bibitem[{{Bongiorno} {et~al.}(2010){Bongiorno}, {Mignoli}, {Zamorani},
  {Lamareille}, {Lanzuisi}, {Miyaji}, {Bolzonella}, {Carollo}, {Contini},
  {Kneib}, {Le F{\`e}vre}, {Lilly}, {Mainieri}, {Renzini}, {Scodeggio},
  {Bardelli}, {Brusa}, {Caputi}, {Civano}, {Coppa}, {Cucciati}, {de la Torre},
  {de Ravel}, {Franzetti}, {Garilli}, {Halliday}, {Hasinger}, {Koekemoer},
  {Iovino}, {Kampczyk}, {Knobel}, {Kova{\v c}}, {Le Borgne}, {Le Brun},
  {Maier}, {Merloni}, {Nair}, {Pello}, {Peng}, {Perez Montero}, {Ricciardelli},
  {Salvato}, {Silverman}, {Tanaka}, {Tasca}, {Tresse}, {Vergani}, {Zucca},
  {Abbas}, {Bottini}, {Cappi}, {Cassata}, {Cimatti}, {Guzzo}, {Leauthaud},
  {Maccagni}, {Marinoni}, {McCracken}, {Memeo}, {Meneux}, {Oesch}, {Porciani},
  {Pozzetti}, \& {Scaramella}}]{2010A&A...510A..56B}
{Bongiorno}, A., {Mignoli}, M., {Zamorani}, G., {et~al.} 2010, \aap, 510, A56+

\bibitem[{{Cid Fernandes} {et~al.}(2011){Cid Fernandes}, {Stasi{\'n}ska},
  {Mateus}, \& {Vale Asari}}]{2011MNRAS.tmp..249C}
{Cid Fernandes}, R., {Stasi{\'n}ska}, G., {Mateus}, A., \& {Vale Asari}, N.
  2011, \mnras, 249

\bibitem[{{Constantin} \& {Vogeley}(2006)}]{2006ApJ...650..727C}
{Constantin}, A. \& {Vogeley}, M.~S. 2006, \apj, 650, 727

\bibitem[{{Contini} {et~al.}(2005){Contini}, {Lemoine-Busserolle}, {Pell{\'o}},
  {Le Borgne}, \& {Kneib}}]{2005RMxAC..24..154C}
{Contini}, T., {Lemoine-Busserolle}, M., {Pell{\'o}}, R., {Le Borgne}, J., \&
  {Kneib}, J. 2005, RMxAC, 24, 154

\bibitem[{{Davis} {et~al.}(2003){Davis}, {Faber}, {Newman}, {Phillips},
  {Ellis}, {Steidel}, {Conselice}, {Coil}, {Finkbeiner}, {Koo}, {Guhathakurta},
  {Weiner}, {Schiavon}, {Willmer}, {Kaiser}, {Luppino}, {Wirth}, {Connolly},
  {Eisenhardt}, {Cooper}, \& {Gerke}}]{2003SPIE.4834..161D}
{Davis}, M., {Faber}, S.~M., {Newman}, J., {et~al.} 2003, SPIE, 4834, 161

\bibitem[{{Garilli} {et~al.}(2008){Garilli}, {Le F{\`e}vre}, {Guzzo},
  {Maccagni}, {Le Brun}, {de la Torre}, {Meneux}, {Tresse}, {Franzetti},
  {Zamorani}, {Zanichelli}, {Gregorini}, {Vergani}, {Bottini}, {Scaramella},
  {Scodeggio}, {Vettolani}, {Adami}, {Arnouts}, {Bardelli}, {Bolzonella},
  {Cappi}, {Charlot}, {Ciliegi}, {Contini}, {Foucaud}, {Gavignaud}, {Ilbert},
  {Iovino}, {Lamareille}, {McCracken}, {Marano}, {Marinoni}, {Mazure},
  {Merighi}, {Paltani}, {Pell{\`o}}, {Pollo}, {Pozzetti}, {Radovich}, {Zucca},
  {Blaizot}, {Bongiorno}, {Cucciati}, {Mellier}, {Moreau}, \&
  {Paioro}}]{2008A&A...486..683G}
{Garilli}, B., {Le F{\`e}vre}, O., {Guzzo}, L., {et~al.} 2008, \aap, 486, 683

\bibitem[{{Garz{\'o}n} {et~al.}(2006){Garz{\'o}n}, {Abreu}, {Barrera},
  {Becerril}, {Cair{\'o}s}, {D{\'{\i}}az}, {Fragoso}, {Gago}, {Grange},
  {Gonz{\'a}lez}, {L{\'o}pez}, {Patr{\'o}n}, {P{\'e}rez}, {Rasilla}, {Redondo},
  {Restrepo}, {Saavedra}, {S{\'a}nchez}, {Tenegi}, \&
  {Vallb{\'e}}}]{2006SPIE.6269E..40G}
{Garz{\'o}n}, F., {Abreu}, D., {Barrera}, S., {et~al.} 2006, SPIE, 6269

\bibitem[{{Groves} {et~al.}(2006){Groves}, {Heckman}, \&
  {Kauffmann}}]{2006MNRAS.371.1559G}
{Groves}, B.~A., {Heckman}, T.~M., \& {Kauffmann}, G. 2006, \mnras, 371, 1559

\bibitem[{{Heckman}(1980)}]{1980A&A....87..152H}
{Heckman}, T.~M. 1980, \aap, 87, 152

\bibitem[{{Ho} {et~al.}(1997){Ho}, {Filippenko}, \&
  {Sargent}}]{1997ApJS..112..315H}
{Ho}, L.~C., {Filippenko}, A.~V., \& {Sargent}, W.~L.~W. 1997, \apjs, 112, 315

\bibitem[{{Kauffmann} {et~al.}(2003{\natexlab{a}}){Kauffmann}, {Heckman},
  {Tremonti}, {Brinchmann}, {Charlot}, {White}, {Ridgway}, {Brinkmann},
  {Fukugita}, {Hall}, {Ivezi{\'c}}, {Richards}, \&
  {Schneider}}]{2003MNRAS.346.1055K}
{Kauffmann}, G., {Heckman}, T.~M., {Tremonti}, C., {et~al.} 2003{\natexlab{a}},
  \mnras, 346, 1055

\bibitem[{{Kauffmann} {et~al.}(2003{\natexlab{b}}){Kauffmann}, {Heckman},
  {White}, {Charlot}, {Tremonti}, {Brinchmann}, {Bruzual}, {Peng}, {Seibert},
  {Bernardi}, {Blanton}, {Brinkmann}, {Castander}, {Cs{\'a}bai}, {Fukugita},
  {Ivezic}, {Munn}, {Nichol}, {Padmanabhan}, {Thakar}, {Weinberg}, \&
  {York}}]{2003MNRAS.341...33K}
{Kauffmann}, G., {Heckman}, T.~M., {White}, S.~D.~M., {et~al.}
  2003{\natexlab{b}}, \mnras, 341, 33

\bibitem[{{Kewley} {et~al.}(2001){Kewley}, {Dopita}, {Sutherland}, {Heisler},
  \& {Trevena}}]{2001ApJ...556..121K}
{Kewley}, L.~J., {Dopita}, M.~A., {Sutherland}, R.~S., {Heisler}, C.~A., \&
  {Trevena}, J. 2001, \apj, 556, 121

\bibitem[{{Kewley} {et~al.}(2006){Kewley}, {Groves}, {Kauffmann}, \&
  {Heckman}}]{2006MNRAS.372..961K}
{Kewley}, L.~J., {Groves}, B., {Kauffmann}, G., \& {Heckman}, T. 2006, \mnras,
  372, 961

\bibitem[{{Lamareille}(2010)}]{2009lama}
{Lamareille}, F. 2010, \aap, 509, A53+

\bibitem[{{Lamareille} {et~al.}(2009){Lamareille}, {Brinchmann}, {Contini},
  {Walcher}, {Charlot}, {P{\'e}rez-Montero}, {Zamorani}, {Pozzetti},
  {Bolzonella}, {Garilli}, {Paltani}, {Bongiorno}, {Le F{\`e}vre}, {Bottini},
  {Le Brun}, {Maccagni}, {Scaramella}, {Scodeggio}, {Tresse}, {Vettolani},
  {Zanichelli}, {Adami}, {Arnouts}, {Bardelli}, {Cappi}, {Ciliegi}, {Foucaud},
  {Franzetti}, {Gavignaud}, {Guzzo}, {Ilbert}, {Iovino}, {McCracken}, {Marano},
  {Marinoni}, {Mazure}, {Meneux}, {Merighi}, {Pell{\`o}}, {Pollo}, {Radovich},
  {Vergani}, {Zucca}, {Romano}, {Grado}, \& {Limatola}}]{2009A&A...495...53L}
{Lamareille}, F., {Brinchmann}, J., {Contini}, T., {et~al.} 2009, \aap, 495, 53

\bibitem[{{Lamareille} {et~al.}(2006{\natexlab{a}}){Lamareille}, {Contini},
  {Brinchmann}, {Le Borgne}, {Charlot}, \& {Richard}}]{2006A&A...448..907L}
{Lamareille}, F., {Contini}, T., {Brinchmann}, J., {et~al.} 2006{\natexlab{a}},
  \aap, 448, 907

\bibitem[{{Lamareille} {et~al.}(2006{\natexlab{b}}){Lamareille}, {Contini}, {Le
  Borgne}, {Brinchmann}, {Charlot}, \& {Richard}}]{2006A&A...448..893L}
{Lamareille}, F., {Contini}, T., {Le Borgne}, J., {et~al.} 2006{\natexlab{b}},
  \aap, 448, 893

\bibitem[{{Lamareille} {et~al.}(2004){Lamareille}, {Mouhcine}, {Contini},
  {Lewis}, \& {Maddox}}]{2004MNRAS.350..396L}
{Lamareille}, F., {Mouhcine}, M., {Contini}, T., {Lewis}, I., \& {Maddox}, S.
  2004, \mnras, 350, 396

\bibitem[{{Le F{\`e}vre} {et~al.}(2010){Le F{\`e}vre}, {Maccagni}, {Paltani},
  {Hill}, {Le Mignant}, {Tresse}, {Garzon Lopez}, {Almaini}, {Brinchmann},
  {Charlot}, {Ciardi}, {Fontana}, {Gallego}, {Garilli}, {Ilbert}, {Meneux}, {de
  Caprio}, {Delabre}, {Genolet}, {Jaquet}, {Martin}, {Roman}, \&
  {Rousset}}]{2010SPIE.7735E..75L}
{Le F{\`e}vre}, O., {Maccagni}, D., {Paltani}, S., {et~al.} 2010, SPIE, 7735

\bibitem[{{Le F{\`e}vre} {et~al.}(2005){Le F{\`e}vre}, {Vettolani}, {Garilli},
  {Tresse}, {Bottini}, {Le Brun}, {Maccagni}, {Picat}, {Scaramella},
  {Scodeggio}, {Zanichelli}, {Adami}, {Arnaboldi}, {Arnouts}, {Bardelli},
  {Bolzonella}, {Cappi}, {Charlot}, {Ciliegi}, {Contini}, {Foucaud},
  {Franzetti}, {Gavignaud}, {Guzzo}, {Ilbert}, {Iovino}, {McCracken}, {Marano},
  {Marinoni}, {Mathez}, {Mazure}, {Meneux}, {Merighi}, {Paltani}, {Pell{\`o}},
  {Pollo}, {Pozzetti}, {Radovich}, {Zamorani}, {Zucca}, {Bondi}, {Bongiorno},
  {Busarello}, {Lamareille}, {Mellier}, {Merluzzi}, {Ripepi}, \&
  {Rizzo}}]{2005A&A...439..845L}
{Le F{\`e}vre}, O., {Vettolani}, G., {Garilli}, B., {et~al.} 2005, \aap, 439,
  845

\bibitem[{{Lilly} {et~al.}(2009){Lilly}, {Le Brun}, {Maier}, {Mainieri},
  {Mignoli}, {Scodeggio}, {Zamorani}, {Carollo}, {Contini}, {Kneib}, {Le
  F{\`e}vre}, {Renzini}, {Bardelli}, {Bolzonella}, {Bongiorno}, {Caputi},
  {Coppa}, {Cucciati}, {de la Torre}, {de Ravel}, {Franzetti}, {Garilli},
  {Iovino}, {Kampczyk}, {Kovac}, {Knobel}, {Lamareille}, {Le Borgne}, {Pello},
  {Peng}, {P{\'e}rez-Montero}, {Ricciardelli}, {Silverman}, {Tanaka}, {Tasca},
  {Tresse}, {Vergani}, {Zucca}, {Ilbert}, {Salvato}, {Oesch}, {Abbas},
  {Bottini}, {Capak}, {Cappi}, {Cassata}, {Cimatti}, {Elvis}, {Fumana},
  {Guzzo}, {Hasinger}, {Koekemoer}, {Leauthaud}, {Maccagni}, {Marinoni},
  {McCracken}, {Memeo}, {Meneux}, {Porciani}, {Pozzetti}, {Sanders},
  {Scaramella}, {Scarlata}, {Scoville}, {Shopbell}, \&
  {Taniguchi}}]{2009ApJS..184..218L}
{Lilly}, S.~J., {Le Brun}, V., {Maier}, C., {et~al.} 2009, \apjs, 184, 218

\bibitem[{{Loubser} {et~al.}(2009){Loubser}, {S{\'a}nchez-Bl{\'a}zquez},
  {Sansom}, \& {Soechting}}]{2009MNRAS.398..133L}
{Loubser}, S.~I., {S{\'a}nchez-Bl{\'a}zquez}, P., {Sansom}, A.~E., \&
  {Soechting}, I.~K. 2009, \mnras, 398, 133

\bibitem[{{Maier} {et~al.}(2009){Maier}, {Lilly}, {Zamorani}, {Scodeggio},
  {Lamareille}, {Contini}, {Sargent}, {Scarlata}, {Oesch}, {Carollo}, {Le
  F{\`e}vre}, {Renzini}, {Kneib}, {Mainieri}, {Bardelli}, {Bolzonella},
  {Bongiorno}, {Caputi}, {Coppa}, {Cucciati}, {de la Torre}, {de Ravel},
  {Franzetti}, {Garilli}, {Iovino}, {Kampczyk}, {Knobel}, {Kova{\v c}}, {Le
  Borgne}, {Le Brun}, {Mignoli}, {Pello}, {Peng}, {Perez Montero},
  {Ricciardelli}, {Silverman}, {Tanaka}, {Tasca}, {Tresse}, {Vergani}, {Zucca},
  {Abbas}, {Bottini}, {Cappi}, {Cassata}, {Cimatti}, {Fumana}, {Guzzo},
  {Halliday}, {Koekemoer}, {Leauthaud}, {Maccagni}, {Marinoni}, {McCracken},
  {Memeo}, {Meneux}, {Porciani}, {Pozzetti}, \&
  {Scaramella}}]{2009ApJ...694.1099M}
{Maier}, C., {Lilly}, S.~J., {Zamorani}, G., {et~al.} 2009, \apj, 694, 1099

\bibitem[{{McLean} {et~al.}(2008){McLean}, {Steidel}, {Matthews}, {Epps}, \&
  {Adkins}}]{2008SPIE.7014E..99M}
{McLean}, I.~S., {Steidel}, C.~C., {Matthews}, K., {Epps}, H., \& {Adkins},
  S.~M. 2008, SPIE, 7014

\bibitem[{{Mouhcine} {et~al.}(2006){Mouhcine}, {Bamford},
  {Arag{\'o}n-Salamanca}, {Nakamura}, \&
  {Milvang-Jensen}}]{2006MNRAS.369..891M}
{Mouhcine}, M., {Bamford}, S.~P., {Arag{\'o}n-Salamanca}, A., {Nakamura}, O.,
  \& {Milvang-Jensen}, B. 2006, \mnras, 369, 891

\bibitem[{{P{\'e}rez-Montero} {et~al.}(2009){P{\'e}rez-Montero}, {Contini},
  {Lamareille}, {Brinchmann}, {Walcher}, {Charlot}, {Bolzonella}, {Pozzetti},
  {Bottini}, {Garilli}, {Le Brun}, {Le F{\`e}vre}, {Maccagni}, {Scaramella},
  {Scodeggio}, {Tresse}, {Vettolani}, {Zanichelli}, {Adami}, {Arnouts},
  {Bardelli}, {Cappi}, {Ciliegi}, {Foucaud}, {Franzetti}, {Gavignaud}, {Guzzo},
  {Ilbert}, {Iovino}, {McCracken}, {Marano}, {Marinoni}, {Mazure}, {Meneux},
  {Merighi}, {Paltani}, {Pell{\`o}}, {Pollo}, {Radovich}, {Vergani},
  {Zamorani}, \& {Zucca}}]{2009A&A...495...73P}
{P{\'e}rez-Montero}, E., {Contini}, T., {Lamareille}, F., {et~al.} 2009, \aap,
  495, 73

\bibitem[{{P{\'e}rez-Montero} {et~al.}(2007){P{\'e}rez-Montero}, {H{\"a}gele},
  {Contini}, \& {D{\'{\i}}az}}]{2007MNRAS.381..125P}
{P{\'e}rez-Montero}, E., {H{\"a}gele}, G.~F., {Contini}, T., \& {D{\'{\i}}az},
  {\'A}.~I. 2007, \mnras, 381, 125

\bibitem[{{Rola} {et~al.}(1997){Rola}, {Terlevich}, \&
  {Terlevich}}]{1997MNRAS.289..419R}
{Rola}, C.~S., {Terlevich}, E., \& {Terlevich}, R.~J. 1997, \mnras, 289, 419

\bibitem[{{Savaglio} {et~al.}(2009){Savaglio}, {Glazebrook}, \& {Le
  Borgne}}]{2009ApJ...691..182S}
{Savaglio}, S., {Glazebrook}, K., \& {Le Borgne}, D. 2009, \apj, 691, 182

\bibitem[{{Sharples} {et~al.}(2006){Sharples}, {Bender}, {Bennett}, {Burch},
  {Carter}, {Casali}, {Clark}, {Content}, {Davies}, {Davies}, {Dubbeldam},
  {Finger}, {Genzel}, {Haefner}, {Hess}, {Kissler-Patig}, {Laidlaw}, {Lehnert},
  {Lewis}, {Moorwood}, {Muschielok}, {F{\"o}rster Schreiber}, {Pirard}, {Ramsay
  Howat}, {Rees}, {Richter}, {Robertson}, {Robson}, {Saglia}, {Tecza},
  {Thatte}, {Todd}, \& {Wegner}}]{2006SPIE.6269E..44S}
{Sharples}, R., {Bender}, R., {Bennett}, R., {et~al.} 2006, SPIE, 6269

\bibitem[{{Stasi{\'n}ska} {et~al.}(2006){Stasi{\'n}ska}, {Cid Fernandes},
  {Mateus}, {Sodr{\'e}}, \& {Asari}}]{2006MNRAS.371..972S}
{Stasi{\'n}ska}, G., {Cid Fernandes}, R., {Mateus}, A., {Sodr{\'e}}, L., \&
  {Asari}, N.~V. 2006, \mnras, 371, 972

\bibitem[{{Stasi{\'n}ska} {et~al.}(2008){Stasi{\'n}ska}, {Vale Asari}, {Cid
  Fernandes}, {Gomes}, {Schlickmann}, {Mateus}, {Schoenell}, \&
  {Sodr{\'e}}}]{2008MNRAS.391L..29S}
{Stasi{\'n}ska}, G., {Vale Asari}, N., {Cid Fernandes}, R., {et~al.} 2008,
  \mnras, 391, L29

\bibitem[{{Tresse} {et~al.}(1996){Tresse}, {Rola}, {Hammer}, {Stasi{\'n}ska},
  {Le Fevre}, {Lilly}, \& {Crampton}}]{1996MNRAS.281..847T}
{Tresse}, L., {Rola}, C., {Hammer}, F., {et~al.} 1996, \mnras, 281, 847

\bibitem[{{Veilleux} \& {Osterbrock}(1987)}]{1987ApJS...63..295V}
{Veilleux}, S. \& {Osterbrock}, D.~E. 1987, \apjs, 63, 295

\bibitem[{{Yan} {et~al.}(2011){Yan}, {Ho}, {Newman}, {Coil}, {Willmer},
  {Laird}, {Georgakakis}, {Aird}, {Barmby}, {Bundy}, {Cooper}, {Davis},
  {Faber}, {Fang}, {Griffith}, {Koekemoer}, {Koo}, {Nandra}, {Park},
  {Sarajedini}, {Weiner}, \& {Willner}}]{2011ApJ...728...38Y}
{Yan}, R., {Ho}, L.~C., {Newman}, J.~A., {et~al.} 2011, \apj, 728, 38

\end{thebibliography}

\end{document}